\documentclass[nofootinbib,superscriptaddress,a4paper,twocolumn,longbibliography,floatfix,aps,prx]{revtex4-2}
\usepackage[english]{babel}
\usepackage[utf8]{inputenc}
\usepackage{amsmath}
\usepackage{amssymb}
\usepackage[colorinlistoftodos, color=green!40, prependcaption]{todonotes}
\usepackage{amsthm}
\usepackage{mathtools}
\usepackage{physics}
\usepackage{xcolor}
\usepackage{graphicx}
\usepackage{adjustbox}
\usepackage{placeins}
\usepackage[T1]{fontenc}
\usepackage{lipsum}
\usepackage{csquotes}
\usepackage[pdftex, pdftitle={Article}, pdfauthor={Author}]{hyperref} 

\usepackage{xcolor}
\hypersetup{
    colorlinks,
    linkcolor={blue!50!black},
    citecolor={blue!50!black},
    urlcolor={blue!50!black}
}

\usepackage{booktabs}

\date{December 10, 2020} 

\begin{document}


\title{Reducing Qubit Requirements while Maintaining Numerical Precision for the Variational Quantum Eigensolver: A Basis-Set-Free Approach}

\author{Jakob~S.~Kottmann}
\email[E-mail:]{jakob.kottmann@utoronto.ca}
\affiliation{Chemical Physics Theory Group, Department of Chemistry, University of Toronto, Canada.}
\affiliation{Department of Computer Science, University of Toronto, Canada.}

\author{Philipp Schleich}
\affiliation{Center for Computational Engineering Science, RWTH Aachen University, Aachen, Germany}

\author{Teresa Tamayo-Mendoza}
\affiliation{Department of Chemistry and Chemical Biology, Harvard University}
\affiliation{Chemical Physics Theory Group, Department of Chemistry, University of Toronto, Canada.}
\affiliation{Department of Computer Science, University of Toronto, Canada.}

\author{Alán Aspuru-Guzik}
\email[E-mail:]{aspuru@utoronto.ca}
\affiliation{Chemical Physics Theory Group, Department of Chemistry, University of Toronto, Canada.}
\affiliation{Department of Computer Science, University of Toronto, Canada.}
\affiliation{Vector Institute for Artificial Intelligence, Toronto, Canada.}
\affiliation{Canadian  Institute  for  Advanced  Research  (CIFAR)  Lebovic  Fellow,  Toronto,  Canada}

\begin{abstract}
    We present a basis-set-free approach to the variational quantum eigensolver using an adaptive representation of the spatial part of molecular wavefunctions. 
    Our approach directly determines system-specific representations of qubit Hamiltonians while fully omitting globally defined basis sets. In this work, we use directly determined pair-natural orbitals on the level of second-order perturbation theory. This results in compact qubit Hamiltonians with high numerical accuracy. We demonstrate initial applications with compact Hamiltonians on up to 22 qubits where conventional representation would for the same systems require 40-100 or more qubits. We further demonstrate reductions in the quantum circuits through the structure of the pair-natural orbitals.
\end{abstract}
\maketitle

Within the framework of the Born{\textendash}Oppenheimer approximation, the electronic structure of molecules is described as a multi-dimensional wave function of electrons in an external potential, usually generated by the charges of the nuclear framework.
Finding accurate approximations to describe those multi-dimensional wave functions is one of the key goals in quantum chemistry, and several types of models have emerged over the last decades.~\cite{aag2018revolution}
Most models formally decompose the multi-dimensional electronic wave function into a linear combination of anti-symmetrized tensor products (Slater determinants) of one-electron wavefunctions (spin-orbitals).
The spin component can be described completely by the two spin-up and spin-down basis states, leaving only the spatial part of the orbitals to be represented.
Traditional electronic structure packages usually use global sets of atom-centered basis functions to describe the spatial parts of the orbitals. Those basis functions mimic the solutions of the Hydrogen atom and the whole procedure is referred to as a linear combination of atomic orbitals (LCAO). The by far most prominent choice are atom-centered Gaussian functions multiplied by polynomial factors.~\cite{fermann2020fundamentals, helgaker2014molecular}
The exponents of the Gaussian functions are globally defined for each atom which leads to a large number of individual basis sets (the EMSL basis set exchange library~\cite{pritchard2019new} lists currently 429 different basis sets for the carbon atom alone), making it a non-trivial task to pick the right basis set for the right computation. Alternatives to Gaussian basis sets, such as exponential functions~\cite{lenthe2003} (Slater-type basis sets) and Sturmians~\cite{herbst2019}, exist and are topic of ongoing research, but are still relying on globally defined basis sets.
Basis-set-free approaches represent the spatial part of the orbitals or other wavefunctions without the use of a globally defined basis set. Some examples are approaches based on Daubechies wavelets~\cite{genovese2008bigdft, mohr2015bigdft,ratcliff2020}, Lagrange-sinc functions~\cite{Kang2020} or multiresolution analysis (MRA)~\cite{harrison2004multiresolution,harrison2016madness, bischoff2019computing, Frediani2013}.
MRA offers an alternative to the traditional basis sets by representing the spatial parts of molecular wavefunctions on adaptive real-space grids, where wavelet-based numerical techniques allow adaptive refinement of the grid in a black-box fashion. In this representation each function (orbital) is described individually by automatically constructed adaptive multi-wavelets making it a basis-set-free representation, since the numerical basis is individually constructed from a proper $L^2$ basis with a numerically well defined truncation criterion. MRA allowed the development of highly accurate quantum chemistry algorithms for ground-state energies~\cite{harrison2004multiresolution, jensen2017elephant}, excitation energies~\cite{kottmann2015numerically, yanai2015multiresolution}, polarizabilities~\cite{brakestad2020, sekino2012new}, magnetic properties~\cite{bischoff2020magnetic, jensen2016magnetic}, as well as relativistic applications~\cite{ANDERSON2020112711, anderson2019dirac} using mean-field and density functional theory (DFT) models.
Initial treatment of correlated methods beyond density functional theory aimed at representing multi-electron wavefunctions directly, resulting in basis-set-free and virtual-orbital-free approaches.~\cite{bischoff2012, bischoff2013, kottmann2017coupledGS, kottmann2017coupledES}
Recently, an approach to directly determine MRA-represented pair natural orbitals (MRA-PNOs) on the level of M\o{}ller-Plesset perturbation theory of second order (MP2) was demonstrated. This approach allows to grow near-optimal system-adapted PNOs from scratch, omitting the use of global basis sets completely. In this work, we will apply those MRA-PNOs in a more general framework beyond the MP2 model, similar to PNO-based methods within LCAO approaches.~\cite{sosa1989selection, deprince2013accurate, riplinger2013, Pinski:2015ii}\\

Classical quantum chemistry algorithms are highly optimized towards Gaussian basis sets (GBS), which makes them the dominant choice of representation.
Quantum algorithms have not reached this highly optimized stage and current research is exploring alternative representations, for example plane-waves~\cite{babbush2019quantum} or Gausslets~\cite{mcclean2020gausslet}, which offer advantages over Gaussian basis sets within the context of quantum computation.
In this work, we introduce a basis-set-free approach to the variational quantum eigensolver (VQE)~\cite{peruzzo2014variational, McClean2016theoryofvqe}, a class of algorithms that variationally minimize the expectation value of a qubit Hamiltonian using a parametrized quantum circuit. The basis-set-free qubit Hamiltonians are constructed from directly determined occupied Hartree{\textendash}Fock orbitals and MRA-PNOs.~\cite{kottmann2020direct} In contrast to global basis sets, the orbitals are optimized system-specific with a surrogate model (MP2), that already accounts for electron correlation. This allows for the construction of compact qubit Hamiltonians, that require a significantly lower number of qubits compared to their LCAO based counterparts.
In variational quantum algorithms, the reduction of the qubit resources has become one of the main objectives, particularly for applications in chemistry. Recent approaches include external corrections, such as the (virtual) quantum subspace expansion~\cite{takeshita_increasing_2020} or explicitly correlated approaches in the form of trans-correlated Hamiltonians~\cite{motta_quantum_2020, mcardle2020improving}, and often come with additional costs in quantum measurements and classical computation.
These methods can as well be applied within the framework of this work and potentially lower the qubit resources even further.
Here, we develop a way to directly construct system-specific compact qubit Hamiltonians with high numerical precision providing a path towards high accuracy quantum chemistry with variational quantum algorithms.

\section{Methodology}
When solving the electronic structure problem, one aims to find approximations for the eigenenergies of electronic Hamiltonians, which for $N_\text{e}$ electrons with coordinates $\Vec{r}_k\in\mathbb{R}^3$ are defined as
\begin{align}
    H\left(\Vec{r}_1, \dots, \Vec{r}_{N_{\text{e}}}\right) = \sum_{k=1}^{N_\text{e}} h\left(\Vec{r}_k\right) + \frac{1}{2}\sum_{k\neq l}^{N_\text{e}} g\left(\Vec{r}_k, \Vec{r}_l\right),
\end{align}
where $h=T+V$ denotes the one-electron kinetic energy operator $T$ with the external potential $V$, and $g(\Vec{r}_k,\Vec{r}_l)$ the electron-electron Coulomb potential. 
For molecules, the external potential $V(\Vec{r})$ is given by the sum of Coulomb potentials between the individual point charges of the nuclei and an electron at position $\Vec{r}$.
The eigenfunctions of the electronic Hamiltonian are anti-symmetric, multi-dimensional functions in $\mathbb{R}^{3N_e}$, making brute-force grid-based computation an unfeasible task.
The requirement of having an anti-symmetrized wavefunction arises from the fermionic nature of the electrons and is usually handled by using Slater determinants or anti-commuting second-quantized operators.\\

In order to tackle this challenge, a large family of approximations to the electronic wavefunction have been introduced. The most prominent is the Hartree{\textendash}Fock (HF) method, that variationally optimizes a single Slater determinant,
therefore reducing the $3N_{\text{e}}$-dimensional problem to $N_{\text{e}}$ coupled, three-dimensional, non-linear problems. 
Improvements upon Hartree{\textendash}Fock, like configuration-interaction (CI) and coupled-cluster (CC) methods, add more anti-symmetrized functions to the wavefunction ansatz, usually created by replacing $n$ orbitals in the initial determinant by correlated $n$-electron functions.
In conventional quantum chemistry, the spatial parts of the Hartree{\textendash}Fock orbitals are represented with globally defined fixed LCAO basis sets for each individual atom.
The Hartree{\textendash}Fock algorithm is then simplified to optimize only the LCAO coefficients, resulting in the Roothaan equations, which often are still referred to as Hartree{\textendash}Fock.
A $N$-orbital basis set results in $2N$-orthonormal spin-orbitals, of which the first $N_{\text{e}}$ define the Hartree{\textendash}Fock reference determinant. This leaves $2N - N_\text{e}$ virtual spin-orbitals free  to represent the correlated electron functions used in CI and CC methods.~\cite{kottmann2017coupledGS}\\

\begin{figure}
    \centering
    \includegraphics[width=0.45\textwidth]{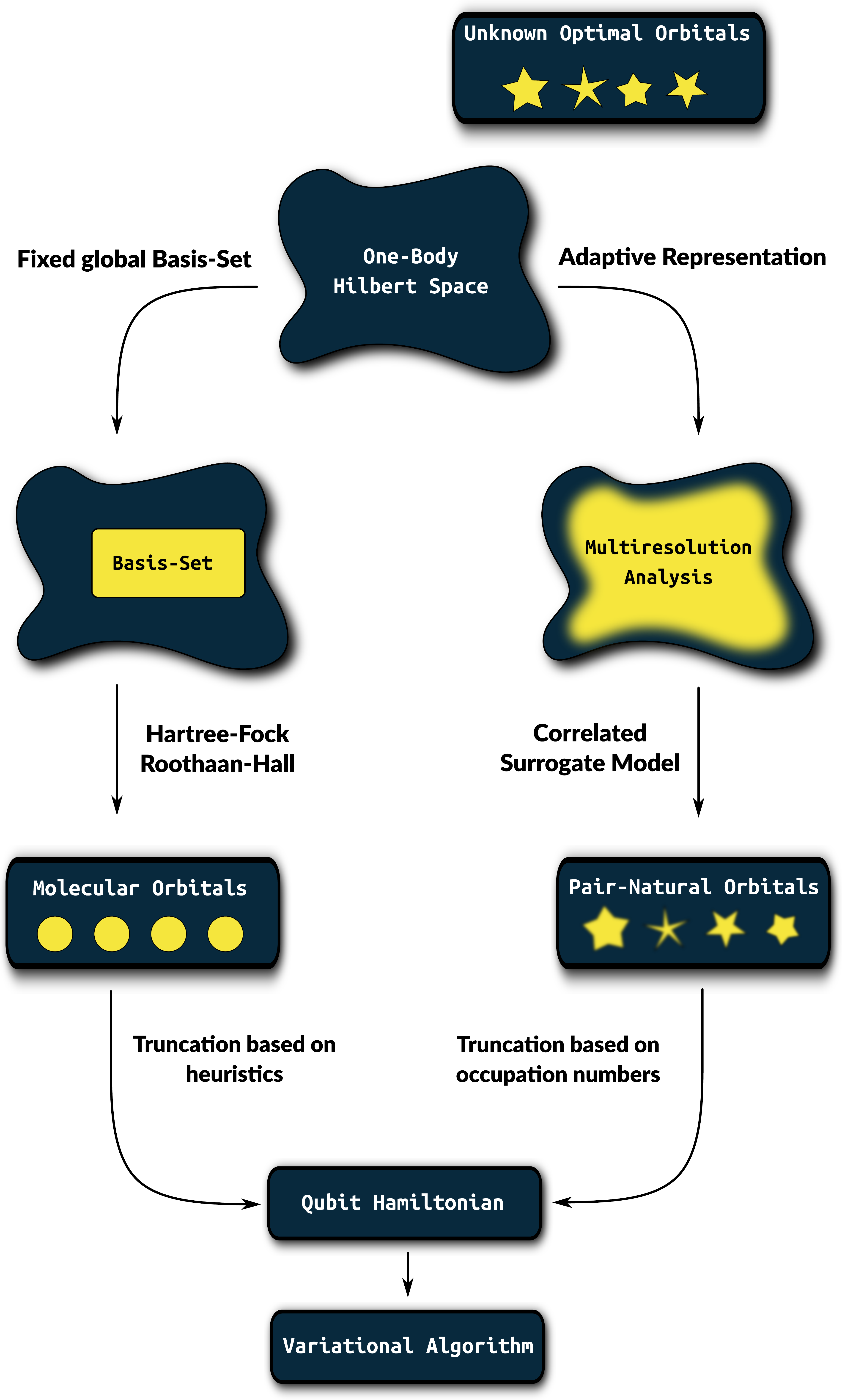}
    \caption{\textbf{Constructing molecular qubit Hamiltonians:} (Left) Representation of the spatial part of molecular orbitals by a fixed set of atom-centered Gaussian functions. This is a global basis set, where functions for each atom are globally defined throughout all possible molecules. (Right) The spatial part of the molecules is represented with multiresolution analysis (MRA) resulting in a locally adaptive representation. Pair natural orbitals (PNOs) are directly determined and optimized within the MRA representation by a surrogate model (in this work MP2). Truncation of the representation to the available qubit number is naturally given by the occupation numbers of the pair-natural orbitalss.}
    \label{fig:high-level-illustration}
\end{figure}

Within the language of second quantization, the description of these methods can be significantly simplified,~\cite{helgaker2014molecular, shavitt2009many, jorgensen2012second, surjan2012second} by expressing the electronic Hamiltonian with abstract field operators~\cite{jordan1927mehrkorperproblem} $\hat{\psi}^\dagger\left(x\right)$, $\hat{\psi}\left(x\right)$, that create or annihilate electron density at spin-coordinate $x=( \Vec{r},\sigma)$
\begin{align}
    H =& \int \operatorname{d}x\; \hat{\psi}^\dagger\left(x\right) h\left(x\right) \hat{\psi}\left(x\right) \label{eq:second_quantized} \\ &+ \frac{1}{2}\int \operatorname{d}x\operatorname{d}y\; \hat{\psi}^\dagger\left(x\right) \hat{\psi}^\dagger\left(y\right) g\left(x,y\right) \hat{\psi}\left(y\right)\hat{\psi}\left(x\right)\nonumber.
\end{align}
Formal expansion of the field operators into an orthogonal set of spin-orbitals as $\hat{\psi}\left(x\right) = \sum_k\varphi_k\left(x\right) \hat{a}_k$ leads to the numerically more suitable form 
\begin{align}
    H &= \sum_{kl} h_{kl} \hat{a}^\dagger_k \hat{a}_l
    +\frac{1}{2}\sum_{klmn} g_{klmn} \hat{a}^\dagger_k \hat{a}^\dagger_l \hat{a}_n \hat{a}_m, \label{eq:second_quantized_finite}
\end{align}
were $h_{kl}$ and $g_{klmn}$ are integrals over spin-orbitals in Dirac notation.
Within the scope of variational quantum algorithms, the second-quantized Hamiltonian can then be transformed to a qubit Hamiltonian using various encodings~\cite{bravyi2002, seeley2012, setia2018}. 
Canonically, when global basis sets are used, the second-quantized Hamiltonian in Eq.~\eqref{eq:second_quantized_finite} is constructed by the occupied and virtual Hartree{\textendash}Fock orbitals. This can be interpreted as pre-optimizing the orbitals within a fixed set of basis functions by a mean-field method.
In this work, we construct the second-quantized Hamiltonian from the occupied Hartree{\textendash}Fock orbitals, solved variationally within a multiresolution analysis representation,~\cite{harrison2004multiresolution} combined with directly determined pair-natural orbitals,~\cite{kottmann2020direct} optimized by MP2. In other words, we are pre-optimizing the orbitals with a correlated method and within a basis-set-free adaptive representation. {The orbitals represented by MRA are basis-set-free since they do not require a globally defined fixed set of basis functions but are instead individually represented by adaptively constructed sets of piecewise polynomials.~\cite{harrison2004multiresolution,harrison2016madness, bischoff2019computing} The construction of the multiresolution representation on a grid is carried out automatically, and for most applications, the user does not require detailed knowledge about the underlying machinery. Hence, from a user perspective, this representation can be treated as \textit{effectively basis-free} and is sometimes referred to as just \textit{basis-free}. To build a Hamiltonian, it is only necessary to set a numerical accuracy threshold (in this work, this is $\epsilon=10^{-4}$) and the required number of orbitals.} In Fig.~\ref{fig:high-level-illustration}, we illustrate the construction of the qubit Hamiltonians using the basis set and the MRA-PNO based basis-set-free approach.
Conceptually, the biggest difference to the canonical construction is, that our approach does not rely on globally fixed sets, but rather optimizes the orbitals directly and system-specific. This allows the freedom to adapt to the molecule at hand, in order to find a close-to-optimal compact representation.
In general, the underlying representation of the pair-natural orbitals can be chosen freely, and does not necessarily need to be MRA. If a sufficient large Gaussian basis-set is chosen, the PNO representation using this basis-set will lead to similar qubit Hamiltonians as obtained with MRA-PNOs. In other words, once computed, the large qubit Hamiltonians of Tab.~\ref{tab:qubit_resources} could be reduced to similar sizes as the qubit Hamiltonians constructed from MRA-PNOs - similar to Ref.~\citenum{barison2020quantum} where such an approach is used with intrinsic atomic orbitals instead of pair natural orbitals. There is however no guarantee, that the chosen Gaussian basis-set is sufficient and picking the right basis-set for the right task depends widely on heuristics and trial and error procedures. We chose MRA-PNOs, since they are usually close to optimally represented pair-natural orbitals, allowing us to focus on the surrogate model itself, without having to speculate about basis-set effects. 
It was shown before, that the direct construction of pair natural orbitals in this way can reach accuracies beyond existing Gaussian basis sets within second-order methods.~\cite{kottmann2020direct} Current implementations can run with molecular systems consisting of dozens of atoms, making it a realistic candidate for future large-scale applications.\\

\subsection{Constructing the Hamiltonian:}
Most MRA-based optimization protocols in quantum chemistry solve a quantum chemical model, defined by the potential $V$, by transforming the Schr\"odinger-like differential equation
\begin{align}
    \left(-\frac{\Delta}{2} + V \right)\ket{\Psi} = E \ket{\Psi}
\end{align}
into an integral equation using the bound-state Helmholtz Green's function $G_E$ as kernel 
\begin{align}
    \Psi\left({r}\right) = -2 \int \operatorname{d}r'\; G_E(r,r') V\left(r'\right) \Psi\left({r'}\right),\label{eq:psi=GVpsi}
\end{align}
leading to an iterative optimization of the wavefunction.~\cite{Kalos1962, beylkin2005alg}
The potential $V$ depends on the underlying quantum chemical model and is usually an effective one-body potential from Hartree{\textendash}Fock and Kohn{\textendash}Sham theory~\cite{harrison2004multiresolution} with corresponding excited state variants~\cite{kottmann2015numerically, yanai2015multiresolution}, or an effective two-body potential from MP2~\cite{bischoff2012, bischoff2013} and Coupled-Cluster approaches~\cite{kottmann2017coupledGS, kottmann2017coupledES}. While more general many-body potentials are possible in theory~\cite{Kottmann2018Coupled-Cluster}, they are computationally challenging and have not been realized yet.
In this work, the potential $V$ is the PNO-MP2 Hylleraas potential, an effective one-body potential described in detail in Ref.~\citenum{kottmann2020direct}. We refer to a recent review~\cite{bischoff2019computing} for further details on MRA-based methods.\\

In this approach, occupied and localized Hartree{\textendash}Fock orbitals are optimized according to Ref.~\citenum{harrison2004multiresolution}, and initial pair-specific guess functions for the PNOs are created by multiplying monomials onto the optimized Hartree{\textendash}Fock orbitals.
These initial PNOs are optimized according to Eq.~\eqref{eq:psi=GVpsi}, where $V$ is determined by the PNO-MP2 Hylleraas functional.~\cite{kottmann2020direct}
{ The optimization results in multiple sets of pair-natural orbitals corresponding to pairs of occupied Hartree{\textendash}Fock orbitals $\ket{i}$ and $\ket{j}$
\begin{align}
	\mathcal{S}_{ij} = \bigcup_{a_{ij}=1}^{r_{ij}} \left\{ \ket{a_{ij}} \right\},\label{eq:orbital_set}
\end{align}
where the ranks $r_{ij}$ depend on the occupation numbers of the pair-natural orbitals, the truncation threshold, and the number of available initial guess functions. In this work, we used two cycles of guess function construction according to Ref.~\citenum{kottmann2020direct}, and set the PNO ranks according to the desired number of qubits.
In order to construct the qubit Hamiltonians, we first globally select the MRA-PNOs $\ket{a_{ij}}$ with the largest occupation numbers and orthonormalize them via Cholesky decomposition. Other orthogonalization methods, like the symmetric Loewdin approach, are also supported. However, we found, that Cholesky behaves better within unitaries that exploit the PNO structure (see later sections), since it preserves the important pair-natural orbitals (ordered by their MP2 occupation number) better. Combining the occupied Hartree{\textendash}Fock orbitals $\ket{i}$,  with the globally orthonormalized pair-natural orbitals $\ket{\tilde{a}_{ij}}$, we obtain an orthonormal set of orbitals
\begin{align}
    \mathcal{S} = \mathcal{S}_\text{HF} \cup \mathcal{S}_\text{PNO}\\
    \mathcal{S}_\text{PNO} = \bigcup_{i\leq j=1}^{N_\text{e}/2} \tilde{\mathcal{S}}_{ij}
\end{align}
where $\mathcal{S}_\text{HF}$ denotes the occupied Hartree{\textendash}Fock orbitals and $\tilde{\mathcal{S}}_{ij}$ the selected and orthonormalized PNOs from Eq.~\eqref{eq:orbital_set}. Due to the orthogonalization the orbitals in $\tilde{\mathcal{S}}_{ij}$ will differ from the original PNOs of the surrogate model.
Finally the second-quantized Hamiltonian is constructed from the orbitals in $\mathcal{S}$.}
This Hamiltonian can be transformed to a qubit Hamiltonian using standard qubit encodings.\\

The computational cost of the MRA-PNO-MP2 surrogate model formally scales as $\mathcal{O}\left(\left(\frac{N_\text{e}}{2}\right)^3 R^2\right)$, where $N_\text{e}$ is the number of electrons and $R$ is bounded from above by the maximal PNO rank $R_\text{max}=\max_{ij} r_{ij}$ and, in practice, behaves like the average rank of the PNO pairs.~\cite{kottmann2020direct} Within this approach, the maximal ranks are fixed and determined by the number of desired qubits $N_\text{q}$ in the qubit Hamiltonian, but are in general not expected to grow asymptotically with system size.~\cite{Neese:2009db, Pinski:2015ii, Werner:2015kw} The formal scaling of the MRA-PNO-MP2 surrogate can then be anticipated to be cubic with system size, which is the same as the natural scaling of MP2 in its real-space formulation.~\cite{kottmann2017coupledGS} The orthonormalization of the PNOs using Cholesky or other techniques can be performed directly in the MRA representation, requiring only the computation of the PNO overlaps , $\mathcal{O}\left(\left(\frac{N_\text{q}}{2}\right)^2\right)$, already required within the MRA-PNO-MP2 calculation, and a transformation of the orbitals, $\mathcal{O}\left(\left(\frac{N_\text{q}}{2}\right)^3\right)$, afterwards. The Hartree{\textendash}Fock calculation within MRA formally scales quadratically with system size and has the potential to be further reduced.~\cite{yanai2004multiresolution, bischoff2019computing}\\

\subsection{Constructing PNO-Specific Unitaries}
{
Since the qubit Hamiltonians of this work are constructed by occupied Hartree{\textendash}Fock orbitals and pair-natural orbitals given in Eq.~\ref{eq:orbital_set}, this additional structure can be exploited in the construction of the quantum circuit. The PNO-MP2 surrogate model itself contains only double excitations from occupied Hartree{\textendash}Fock orbitals $\ket{i},\ket{j}$ to the pair-specific PNOs $\ket{a_{ij}}$, and this excitation structure can be transformed directly to construct unitary quantum circuits.
In the case of pair-excitation models, such as the $k$-UpCCGSD model, we can define the PNO-restriced doubles and generalized doubles models as
\begin{align}
    &\ket{\text{PNO-UpCCD}} = U_{\tilde{\text{D}}}U_{\text{HF}}\ket{0},\label{eq:def_pno-UpCCD}\\
    &\ket{\text{PNO-UpCCGD}} = U_{\text{G}\tilde{\text{D}}}U_{\tilde{\text{D}}}U_{\text{HF}}\ket{0},\label{eq:def_pno-UpCCGD}
\end{align}
using the PNO-restricted unitary operators
\begin{align}
    &U_{\tilde{\text{D}}} = \prod_{i=1}^{\frac{N_\text{e}}{2}}\prod_{a\in\mathcal{\tilde{S}}_{ii}} e^{-i\frac{\theta}{2} \tilde{G}_{i a i a}}\\
    &U_{\text{G}\tilde{\text{D}}}=\prod_{i=1}^{\frac{N_\text{e}}{2}}\prod_{a,b\in\mathcal{\tilde{S}}_{ii}} e^{-i\frac{\theta}{2} \tilde{G}_{a b a b}}
\end{align}
and the pair excitation generator defined as
\begin{align}
    \tilde{G}_{i a i a} = i\left( a^\dagger_{a_\uparrow}a_{i_\uparrow}a^\dagger_{a_\downarrow}a_{i_\downarrow} - h.c. \right).
\end{align}
The incorporation of (generalized) singles can be realized in the same way, with the single excitation generators corresponding to the used double excitation generators.
The PNO-UpCCD wavefunction has a product structure of electron-pair wavefunctions, that themselves are supported by Hartree{\textendash}Fock orbitals $\ket{i}$ and the corresponding PNOs $\ket{a_{ii}}$.
The doubles-only variants (PNO-UpCCD and PNO-UpCCGD) of this low-depth approaches offer additional advantages. Due to the restriction to pair-excitations resulting from the same spatial orbitals, these wavefunctions allow direct encoding~\cite{elfving2020simulating,khamoshi2020correlating} of spatial orbitals into qubits, reducing the qubit requirements by a factor of two. For example, for the LiH molecule this is possible without notable loss of accuracy (see Fig.~\ref{fig:pes_plots}). In addition, the model depends only on diagonal pairs $\ket{a_{ii}}$, lowering the computational demands of the employed surrogate model. Detailed strategies in order to exploit the product structure of the wavefunction in quantum and classical simulations are currently investigated. Note, that those additional reductions are not denoted in Tab.~\ref{tab:qubit_resources}.
}\\

\section{Initial Applications}
To evaluate the accuracy and efficiency of the proposed approach, we employ MRA-PNO-MP2 as a basis-set-free surrogate model to the variational quantum eigensolver. We employed the UpCCGSD model of Ref.~\citenum{lee2018generalized} to construct the quantum circuits.
All used model systems are chosen such that they are well described by this ansatz, which allows to focus on the numerical accuracy of the qubit Hamiltonians without worrying about the quality of the ansatz.
An overview of the qubit requirements using MRA and GBS representations is given in Tab.~\ref{tab:qubit_resources}, where we report significant improvement for all systems and types of energy metrics investigated in this work.
As energy metrics we used non-parallelity (NPE) and maximum (MAX) errors and a reaction barrier. Non-parallelity errors~\cite{lee2018generalized} are defined as the difference between the maximal and minimal error on a given potential energy surface. Note that, other than in Ref.~\citenum{lee2018generalized}, the reference values are here also chosen with respect to the underlying one-particle basis.
The MRA-PNOs are optimized according to Ref.~\citenum{kottmann2020direct} using \textsc{madness}~\cite{harrison2016madness}.
Note, that in this work the MRA-PNOs were optimized without regularizing the Coulomb singularity.\\

VQE calculations are performed with \textsc{tequila}~\cite{tequila} using \textsc{qulacs}~\cite{qulacs} as quantum backend, the BFGS optimizer of \textsc{scipy}~\cite{scipy} and the qubit encodings of \textsc{openfermion}~\cite{openfermion}.
Analytical gradients for the BFGS optimization were obtained automatically through the techniques described in Ref.~\citenum{kottmann2020feasible}.
LCAO reference calculations are performed with \textsc{psi4}~\cite{psi42020}.
In all VQE calculations the parameters are initialized as zero - \textit{i.e.} starting from the Hartree{\textendash}Fock reference state.
Representations of Hamiltonians are abbreviated with MRA($N_\text{e}$,$N_\text{q}$) for MRA-PNOs and the acronym for standard LCAO basis sets. The values in parentheses represent the number of electrons $N_\text{e}$ and qubits (spin-orbitals) $N_\text{q}$ .
Note, that classical FCI calculations with basis sets corresponding to large qubit Hamiltonians with 50 or more qubits are possible, since these algorithms are not operating in the full Fock space of the second quantized Hamiltonian.
For simplicity, we omitted known general compression schemes that allow to reduce the number of qubits by two when combined with parity based encodings~\cite{bravyi2017tapering}, since these would apply to all qubit Hamiltonians in this work in the same way. The numerical accuracy of the qubit Hamiltonian is independent of the encoding and the results of this work were obtained with the Jordan-Wigner representation, our implementation within \textsc{tequila} however does support other encodings. The qubit encoding can influence the results of possible future demonstrations on real quantum hardware, since it will result in different gate decomposition of the VQE unitary and therefore will have varying properties with regard to the specifics of the device noise characteristics.

\begin{table}
    \begin{tabular}{ccccc}
    \toprule
    System & Metric & MRA & GBS & More  \\
    \midrule
    He & MAX & \textbf{4} & 4-10 & Fig.~\ref{fig:atoms} \\
    \midrule
    Be & MAX & \textbf{10} & 10-18 & Fig.~\ref{fig:atoms} \\
    \midrule
         H$_2$ & NPE & \textbf{4} & 20-56 & Figs.~\ref{fig:pes_plots},~\ref{fig:pes_npe_max} \\
         H$_2$ & NPE & \textbf{8} & 20-56  & Figs.~\ref{fig:pes_plots},~\ref{fig:pes_npe_max} \\
         H$_2$ & NPE & \textbf{20} & 56-120& Figs.~\ref{fig:pes_plots},~\ref{fig:pes_npe_max} \\
         H$_2$ & MAX & \textbf{4} & 8      & Figs.~\ref{fig:pes_plots},~\ref{fig:pes_npe_max} \\
         H$_2$ & MAX & \textbf{8} & 20-56  & Figs.~\ref{fig:pes_plots},~\ref{fig:pes_npe_max} \\
         H$_2$ & MAX & \textbf{20} & 56    & Figs.~\ref{fig:pes_plots},~\ref{fig:pes_npe_max} \\
         \midrule
         LiH & NPE & \textbf{12-22} & 38-88  & Figs.~\ref{fig:pes_plots},~\ref{fig:pes_npe_max} \\
         LiH & MAX & \textbf{12} & 38-88  & Figs.~\ref{fig:pes_plots},~\ref{fig:pes_npe_max} \\
         LiH & MAX & \textbf{22} & 170-288 & Figs.~\ref{fig:pes_plots},~\ref{fig:pes_npe_max} \\
         \midrule
         BH & NPE & \textbf{12-22} & 38-88  & Figs.~\ref{fig:pes_plots},~\ref{fig:pes_npe_max} \\
         BH & MAX & \textbf{12-22} & 38-88 & Figs.~\ref{fig:pes_plots},~\ref{fig:pes_npe_max} \\
         \midrule
         BeH$_2$ & NPE & \textbf{12} & 46-114 & Figs.~\ref{fig:pes_plots},~\ref{fig:pes_npe_max} \\
         BeH$_2$ & MAX & \textbf{12} & 24-46 & Figs.~\ref{fig:pes_plots},~\ref{fig:pes_npe_max}\\
    \midrule
    NH$_3$ & $\Delta$E & \textbf{12-18}  & 58-100 & Fig.~\ref{fig:nh3_barrier} \\
    \end{tabular}
        \caption{\textbf{Qubit requirements of MRA and GBS representations:} Qubit requirements for the MRA Hamiltonians used in this work compared to qubit requirements using standard Dunning-type basis sets (GBS) that achieve comparable accuracy within different metrics. The employed metrics are non-parallelity (NPE) and maximum (MAX) errors on potential energy surfaces and a reaction barrier ($\Delta E$).}
    
    \label{tab:qubit_resources}
\end{table}

\begin{table}[]
    \centering
    \begin{tabular}{cccc}
    \toprule
    & A & B & C \\
    \midrule
    LiH(4,12) & 4 (192) & 12 (352) & 45 (1280) \\
    BH(6,12) & 3 (144) & 9 (240) & 45 (1280) \\ 
    BeH$_2$(4,12) & 4 (192) & 12 (368) & 45 (1280) \\
    \midrule
    LiH(4,22) & 9 (432) & 27 (1216) & 165 (6160) \\
    BH(6,22) & 7 (336) & 21 (848) & 165 (6160)\\
    \bottomrule
    \end{tabular}
    \caption{\textbf{Parameter and CNOT counts:} Number of parameters (fermionic excitations) for PNO-UpCCD~(A), PNO-UpCCSD~(B), UpCCGSD~(C). See Fig.~\ref{fig:pes_plots} for the corresponding energies. Numbers in parentheses denote CNOT counts from a naive, non-optimized decomposition of the unitaries as a first estimate. More advanced circuit construction is anticipated to significantly reduce those counts.}
    \label{tab:parameter_counts}
\end{table}

\subsection{Helium and Beryllium Atom}
\begin{figure*}
    \begin{tabular}{cc}
    A & B \\
    \includegraphics[width=0.49\textwidth]{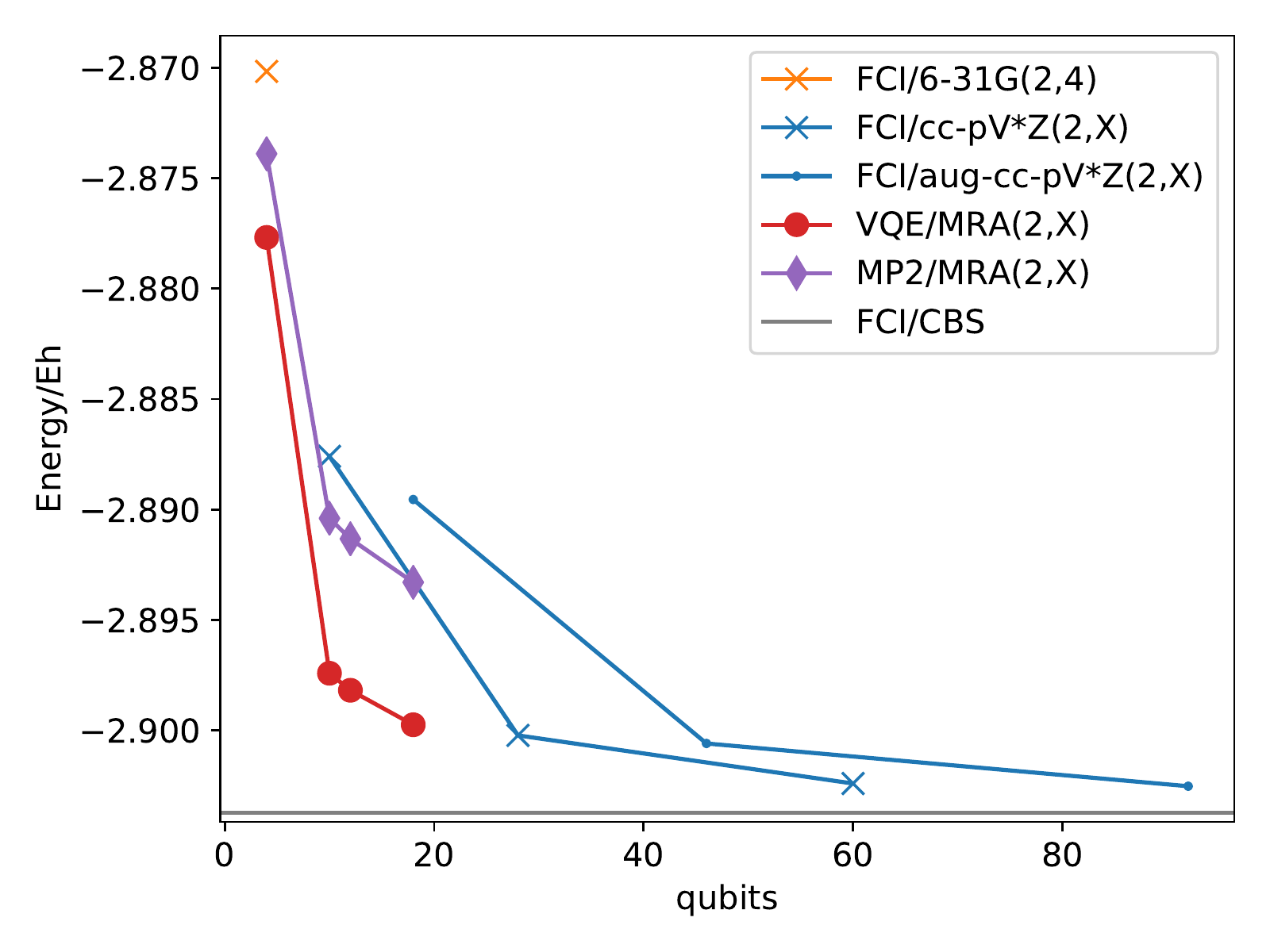}&
    \includegraphics[width=0.49\textwidth]{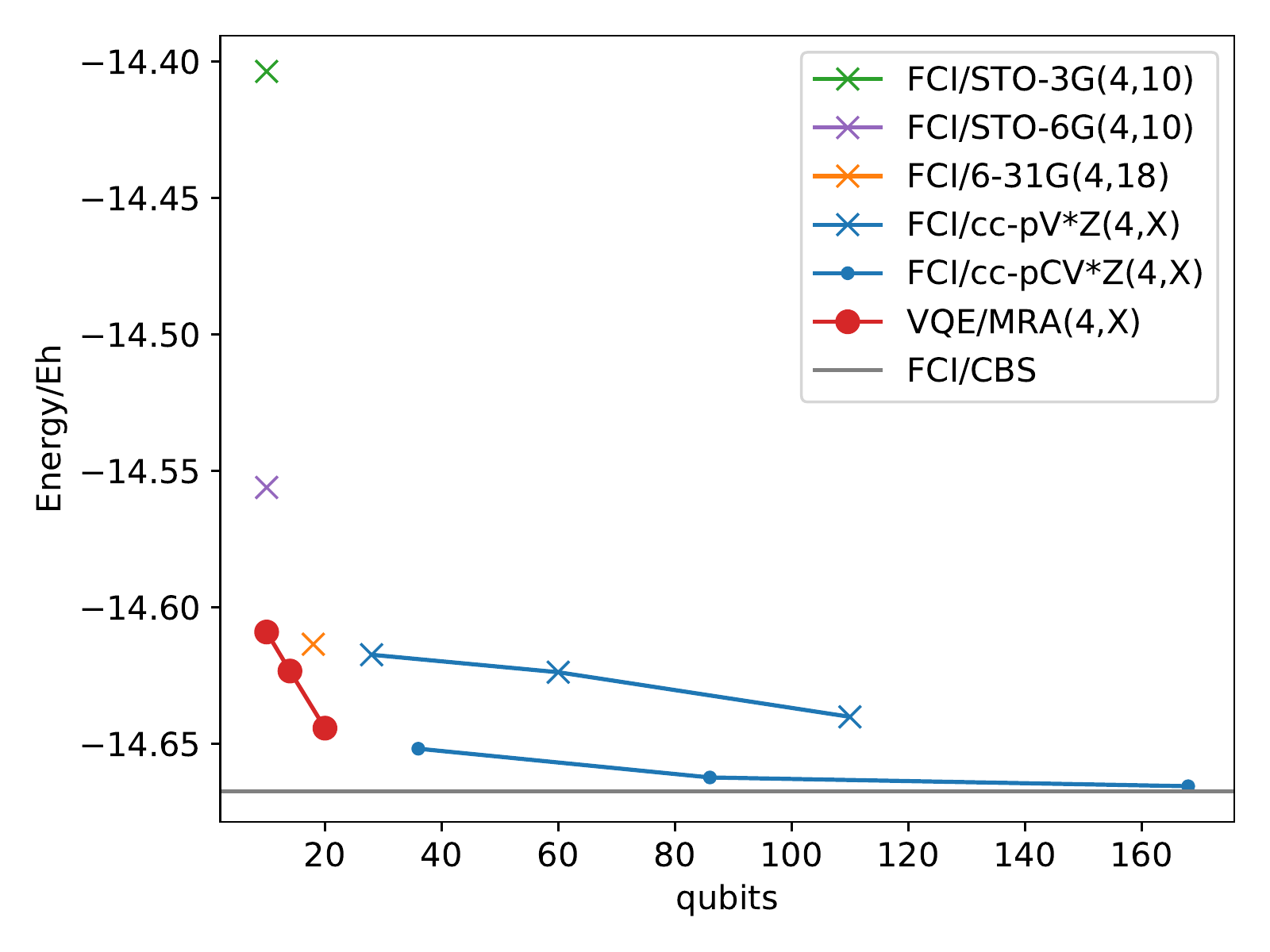}\\
    \end{tabular}
    \caption{\textbf{Helium (A) and beryllium (B) atoms:} Classical full-CI energies compared with basis-set-free VQE energies for the Helium (A) and Beryllium (B) atoms. The VQE part uses the UpCCGSD ansatz. UpCCGSD results with small Gaussian basis sets are nearly indistinguishable from the FCI results and are omitted in the plots. The reference values FCI/CBS are $-2.9037$~\cite{dehesa_study_1992, hiroshi2012, bischoff2014regularizingmany} and $-14.667$~\cite{hiroshi2012} in Hartree units. }
    \label{fig:atoms}
\end{figure*}

As an initial test, we computed the energies of the Helium and Beryllium atom, where the true energies close to the complete basis set (CBS) limit are known from different highly accurate numerical calculations in the literature.~\cite{dehesa_study_1992,hiroshi2012,bischoff2014regularizingmany}
The results are shown in Fig.~\ref{fig:atoms}, where we show the absolute energies with respect to the qubit requirements. MRA-MP2 results are shown for the Helium atom to illustrate that the problem is not fully solved by the surrogate model.
For atoms, LCAO basis sets are expected to perform well, and the basis-set-free Hartree{\textendash}Fock orbitals and PNOs have the same shape as atomic orbitals. One of the main differences is the missing \textit{nuclear cusp}~\cite{Kato1957cusp, bischoff2014regularizing}, present in all molecular wavefunctions due to the singularities in the nuclear Coulomb potential, and non-representable by primitive Gaussian functions. Gaussian basis sets mitigate this through contracted Gaussian basis functions. The \textit{STO-3G} basis set for example uses three Gaussians to represent one contracted basis function while the \textit{STO-6G} uses six, explaining the significantly better performance of \textit{STO-6G} compared to \textit{STO-3G} in Fig.~\ref{fig:atoms}.
The other basis sets are expected to represent the nuclear cusp well. Their slow convergence is a known phenomenon resulting from the \textit{electronic cusp}~\cite{Kato1957cusp, kutzelnigg1985r} in the wavefunctions, generated by the singularities in the electron-electron potential and, in general, hard to represent in a separated representation. So, this is also the case for the MRA-PNOs in this work, since the many-electron wavefunction is still represented as a separated representation using one-electron wavefunctions.  For the Helium atom, a decrease in the slope of the convergence can already be observed, and we expect the same for the Beryllium atom with increasing number of qubits. The basis-set-free approach however achieves significantly more accurate results compared to the LCAO representations. In future works, the representation could be further improved by explicitly correlated approaches that correct the inaccurately description of the electronic cusp.~\cite{bischoff2014regularizingmany} These techniques can already be applied in the MRA-PNO optimization~\cite{kottmann2020direct}, and recently such methods have been introduced for VQEs in the form of transcorrelated Hamiltonians.~\cite{motta_quantum_2020, mcardle2020improving}

\begin{figure*}
    \includegraphics[width=0.95\textwidth]{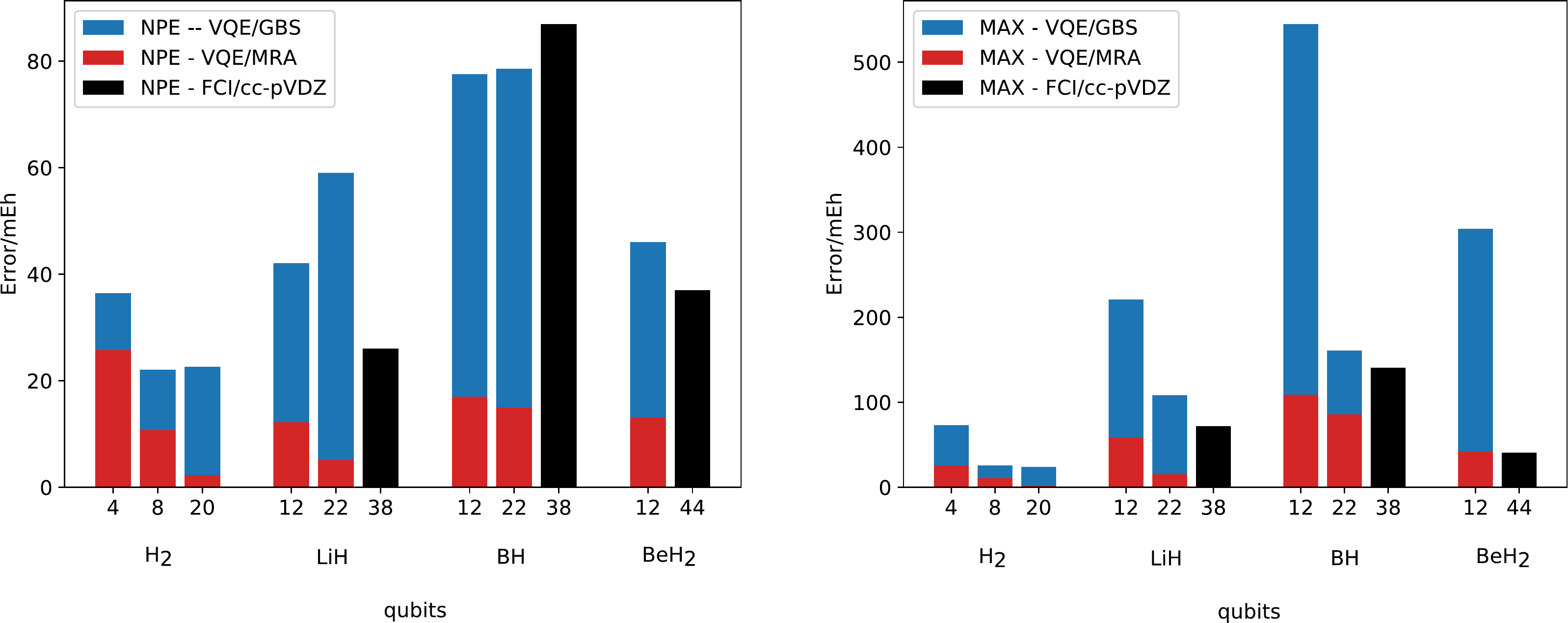}
    \caption{\textbf{Non-parallelity (NPE) and maximum errors (MAX):} Comparison of NPE and MAX errors of the potential energy surfaces of Fig.~\ref{fig:pes_plots} using standard Gaussian basis sets (GBS) or the basis-set-free VQE approach (MRA). The used reference values are FCI/cc-pVQZ(2,120) for H$_2$, FCI/cc-pCVQZ(4,228) for LiH, CISDTQ/cc-pCVTZ(6,114) for BH and FCI/cc-pVQZ(4,224) for BeH$_2$. The GBS for the VQE are STO-3G, 6-31G and cc-pVDZ.}
    \label{fig:pes_npe_max}
\end{figure*}

\begin{figure*}
    \begin{tabular}{cc}
    \includegraphics[width=0.49\textwidth]{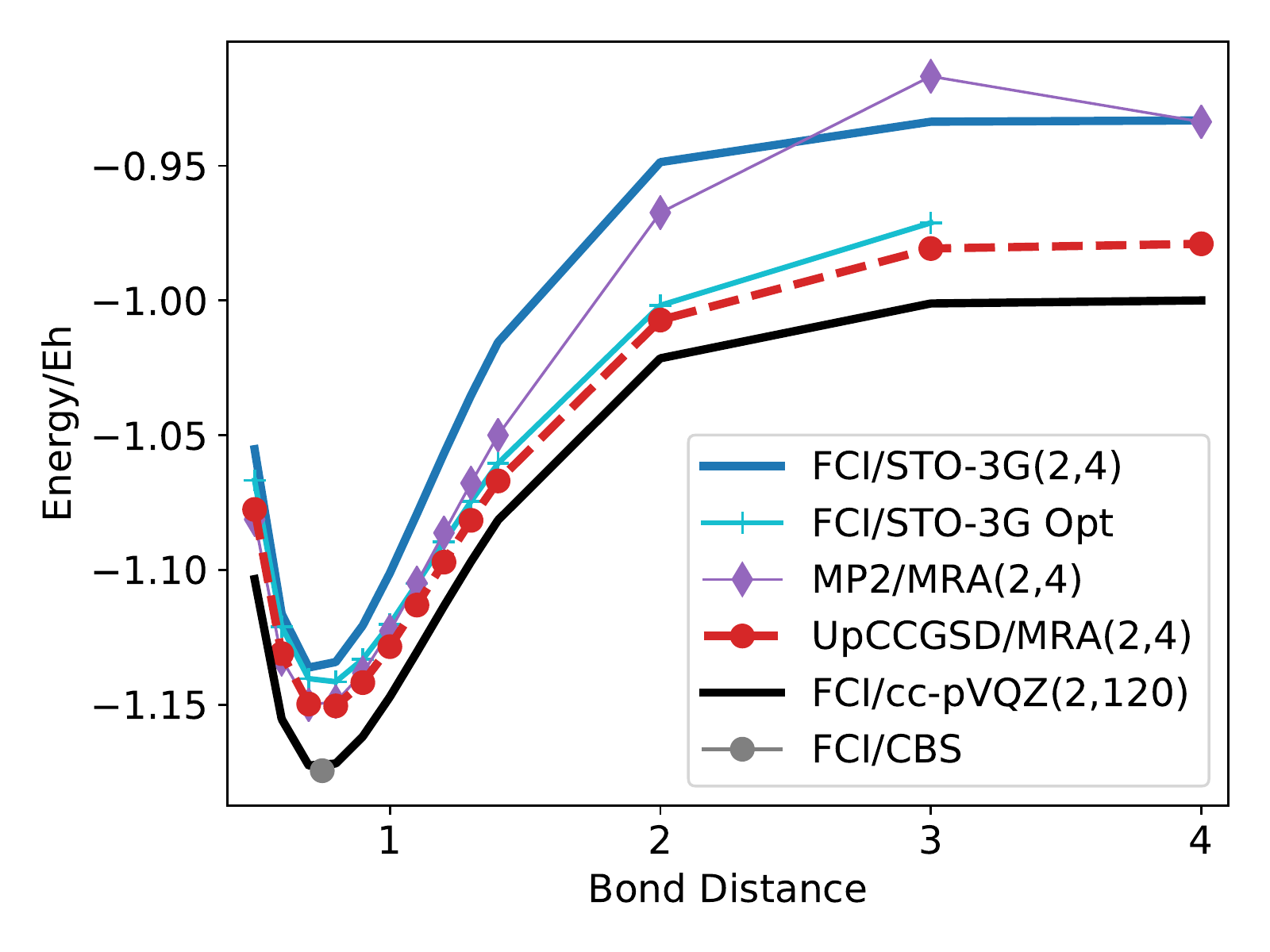}&
    \includegraphics[width=0.49\textwidth]{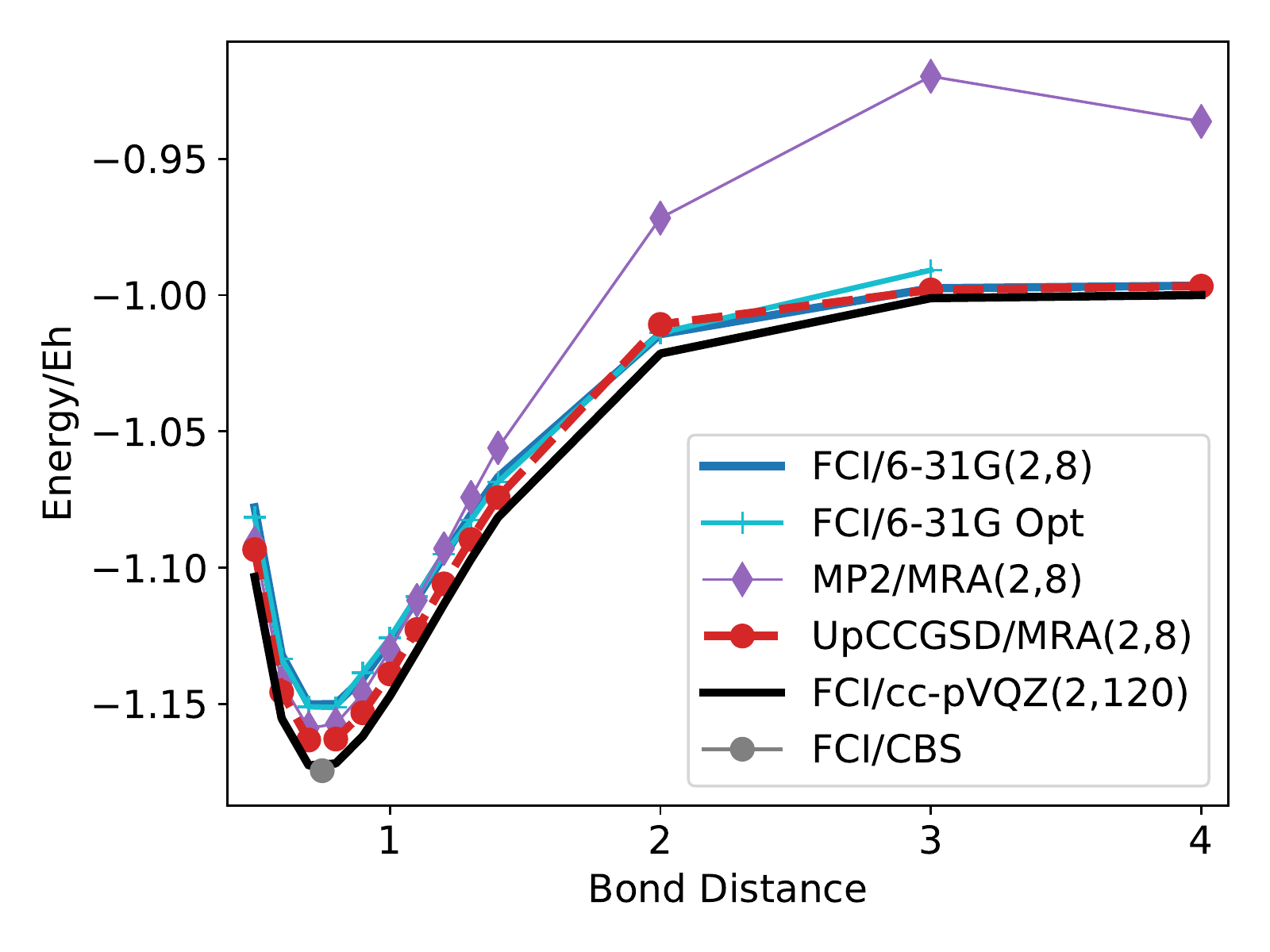}
    \end{tabular}
    \caption{\textbf{Dissociation curve of the hydrogen molecule:} Energies are computed with a VQE (UpCCGSD) or classical methods (FCI). The Hamiltonians were constructed with basis-set-free methods (MRA) and Gaussian basis sets (GBS). All curves are denoted as method/basis($N_\text{e}$, $N_\text{q}$) with number of electrons $N_\text{e}$ and number of qubits $N_\text{q}$. Blue curves show GBS results with minimal (STO-3G) and slightly larger (6-31G) basis sets, while red curves show the UpCCGSD/MRA results with the same qubit requirements. Black curves show the best GBS results (the same in both panels).
    The two H$_2$/GBS-Opt curves show results for the individually optimized Gaussian basis sets (exponents, contraction coefficients, centers) according to Ref.~\citenum{tamayo2018}.
    The MP2/MRA($N_\text{e}$, $N_\text{q}$) curves show the result of the classical surrogate model (PNO-MP2) for the Hydrogen molecule. The same orbitals are used for UpCCGSD/MRA($N_\text{e}$, $N_\text{q}$), illustrating the energy differences between the surrogate model and the VQE. FCI/CBS points were taken from Refs.~\citenum{bande_lih_2010,hiroshi2012}. Bond distances are given in {\AA}ngstrom. }
    \label{fig:h2_plot}
\end{figure*}

\subsection{Bond Dissociation Curves}
\begin{figure*}
    \centering
    \begin{tabular}{ccc}
    \toprule
        \multicolumn{3}{c}{LiH}\\
    \midrule
    \includegraphics[width=0.32\textwidth]{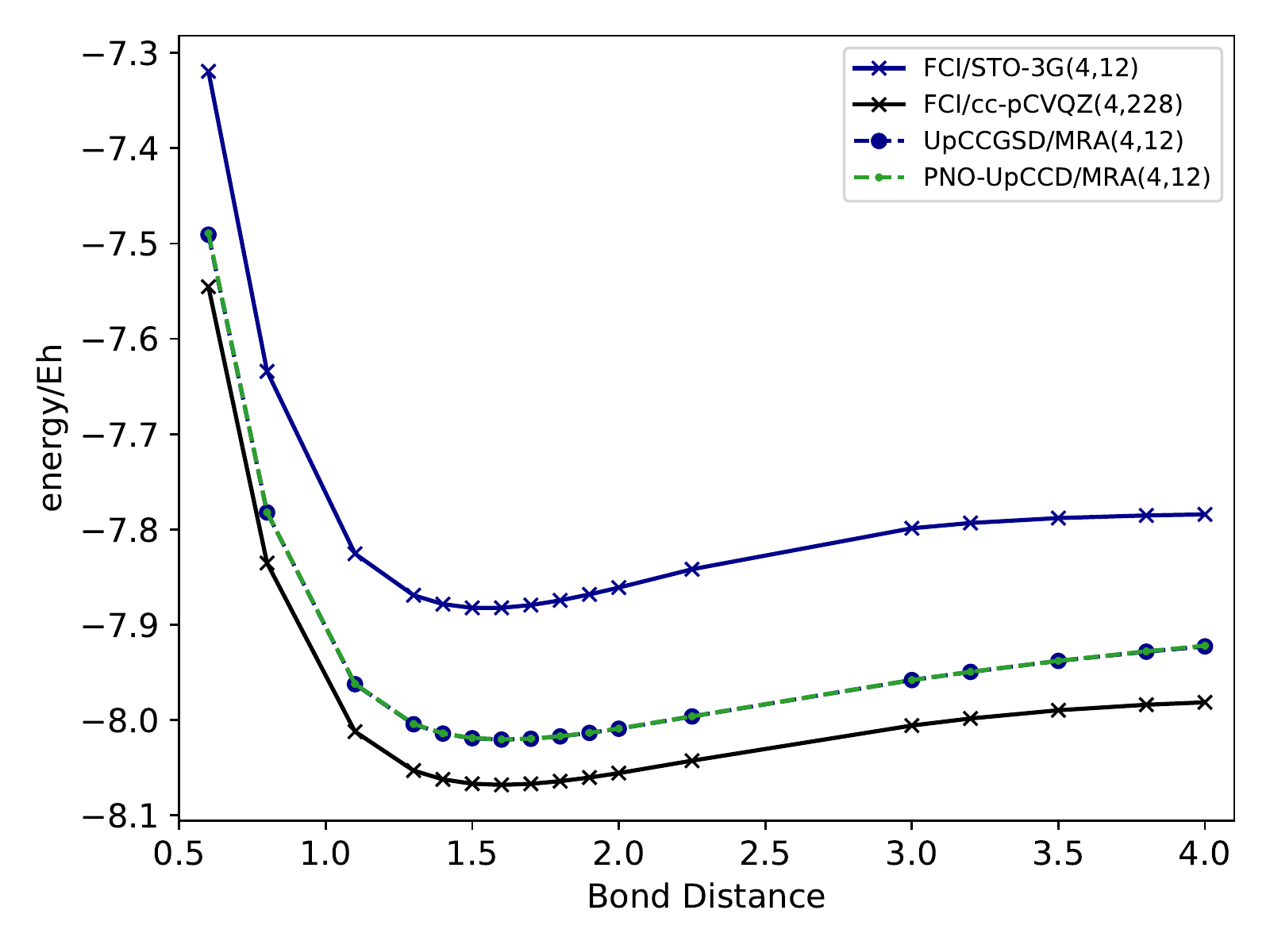}&
    \includegraphics[width=0.32\textwidth]{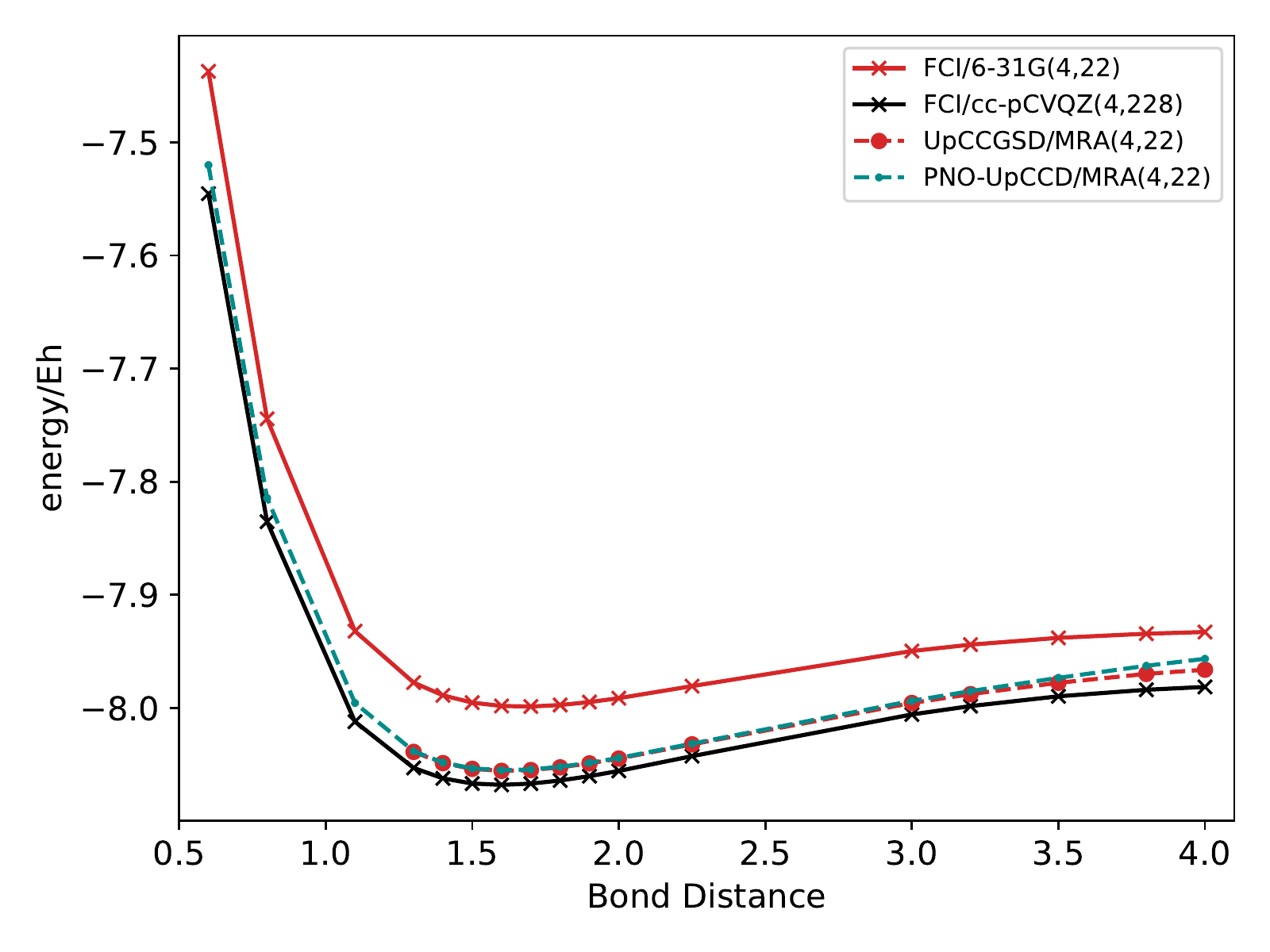}&
    \includegraphics[width=0.32\textwidth]{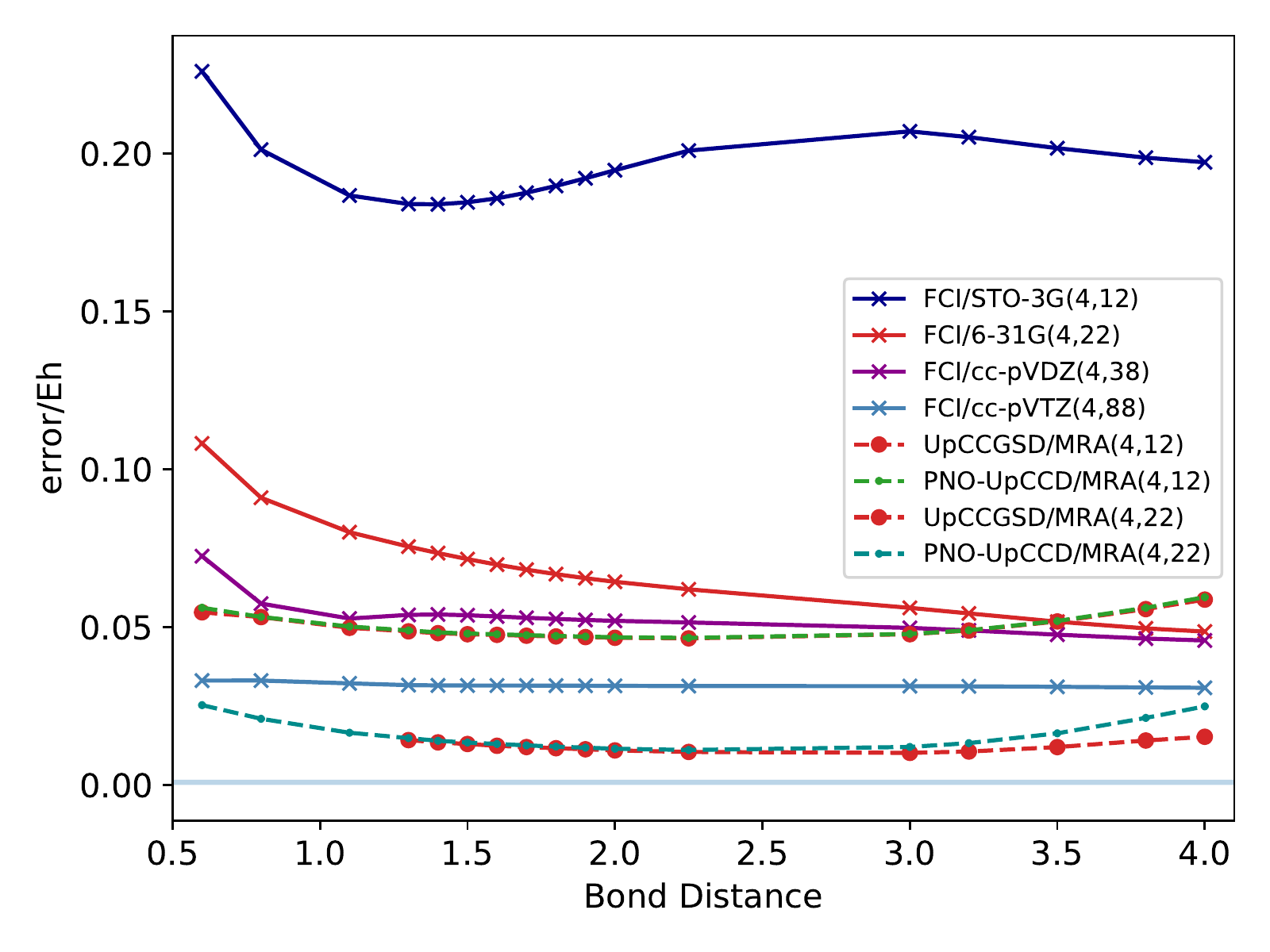}\\
    \midrule
        \multicolumn{3}{c}{BH}\\
    \midrule
    \includegraphics[width=0.32\textwidth]{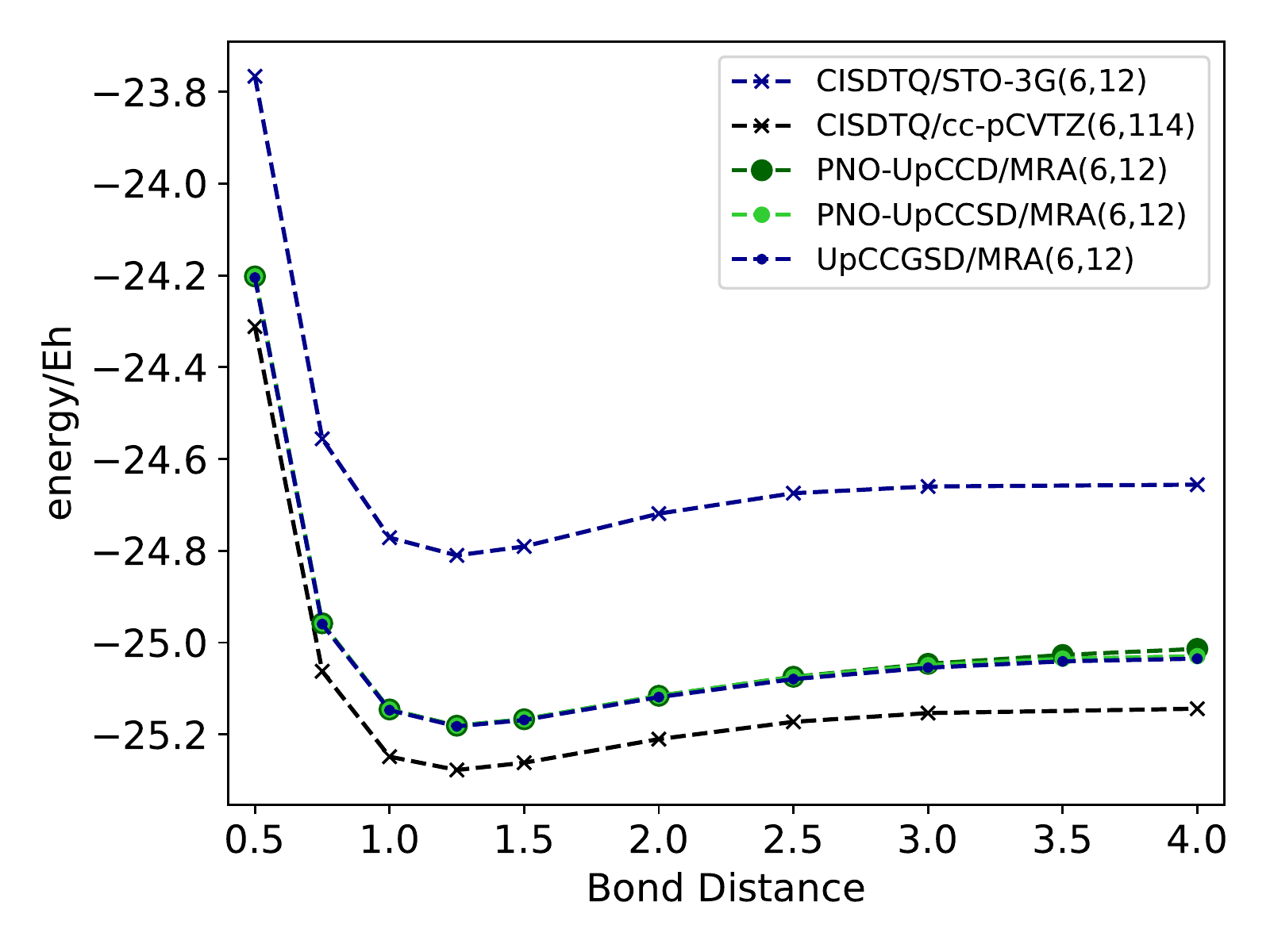}&
    \includegraphics[width=0.32\textwidth]{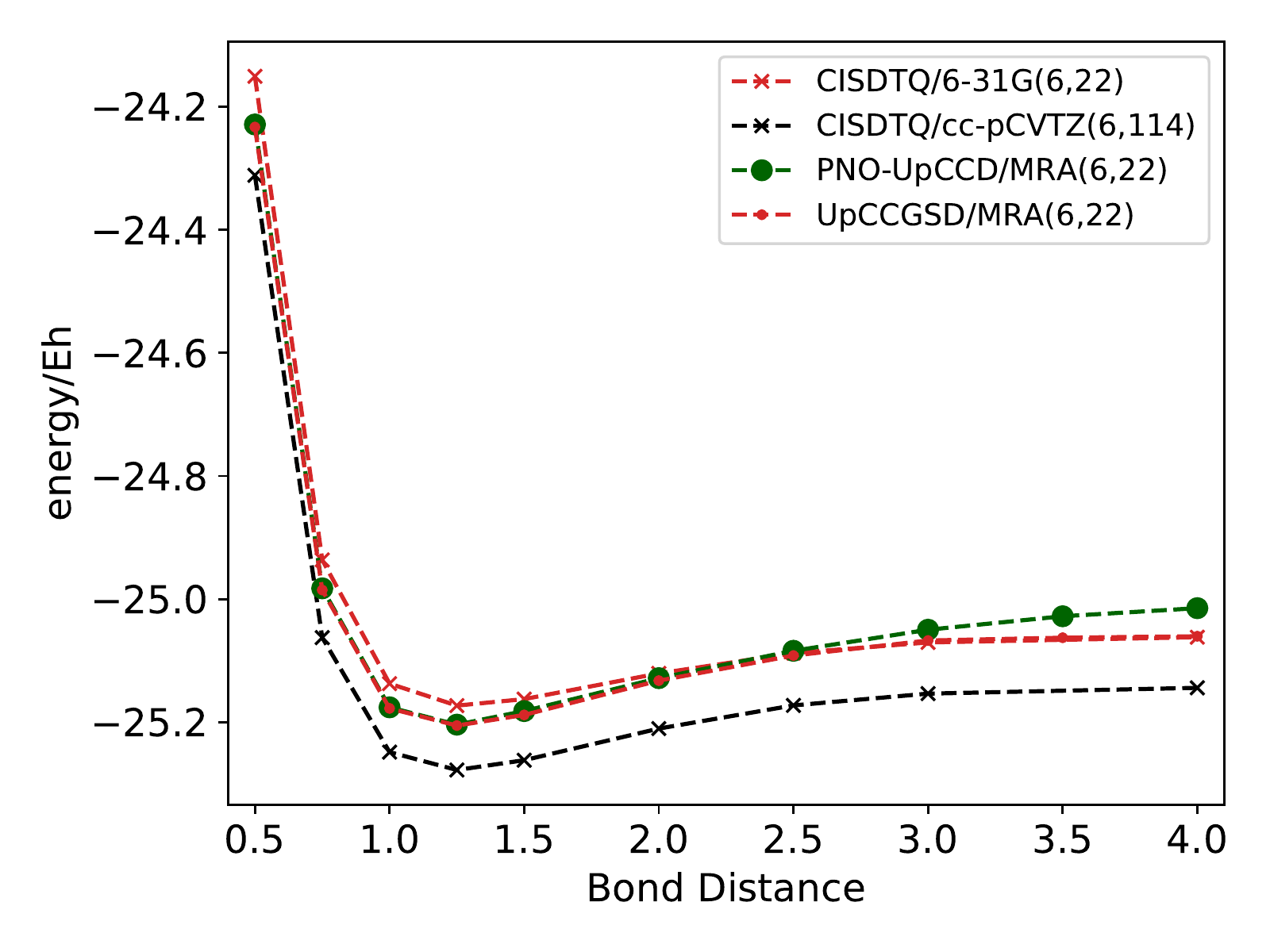}&
    \includegraphics[width=0.32\textwidth]{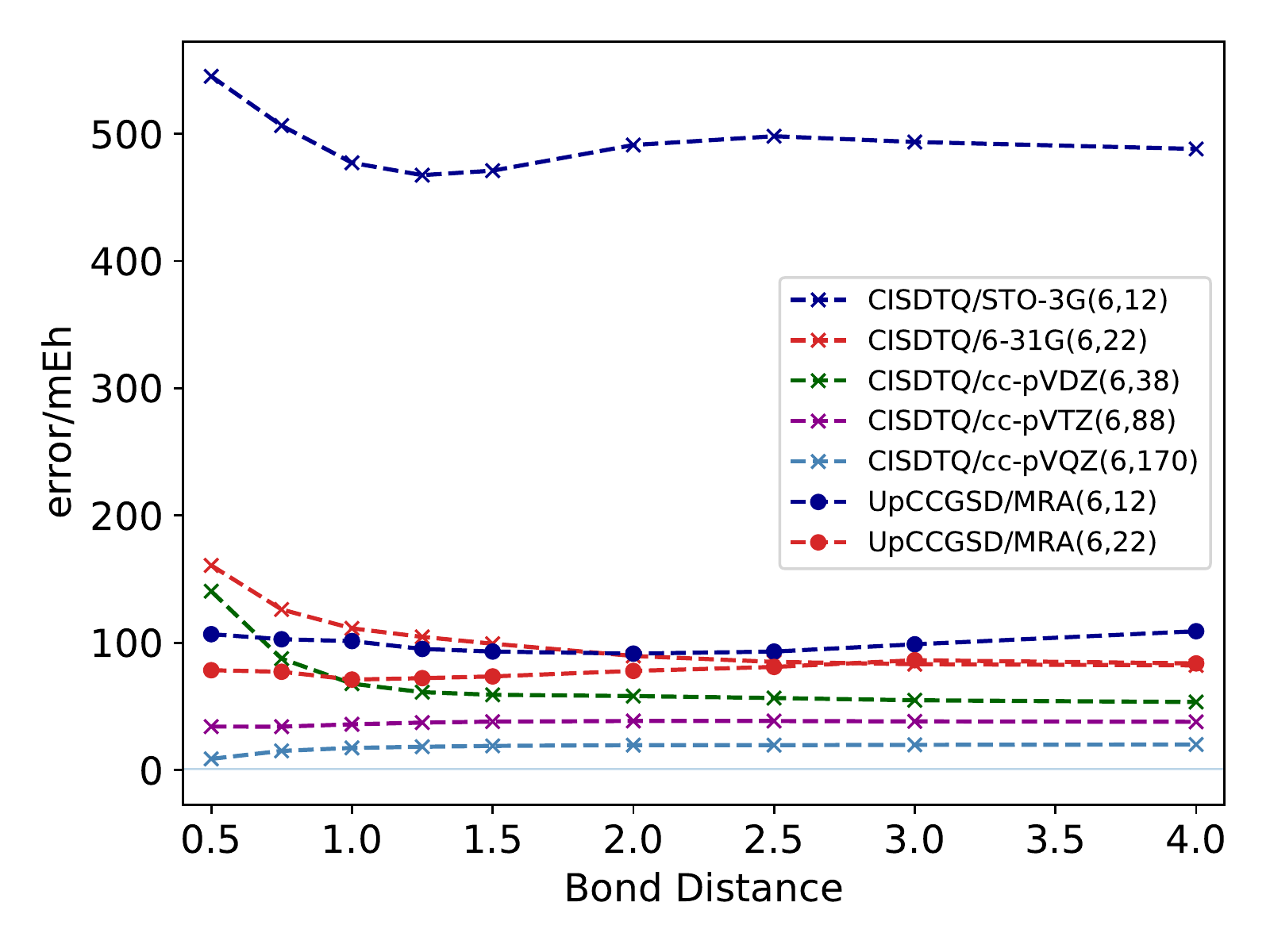}\\
    \midrule
        \multicolumn{3}{c}{BeH$_2$}\\
    \midrule
    \includegraphics[width=0.32\textwidth]{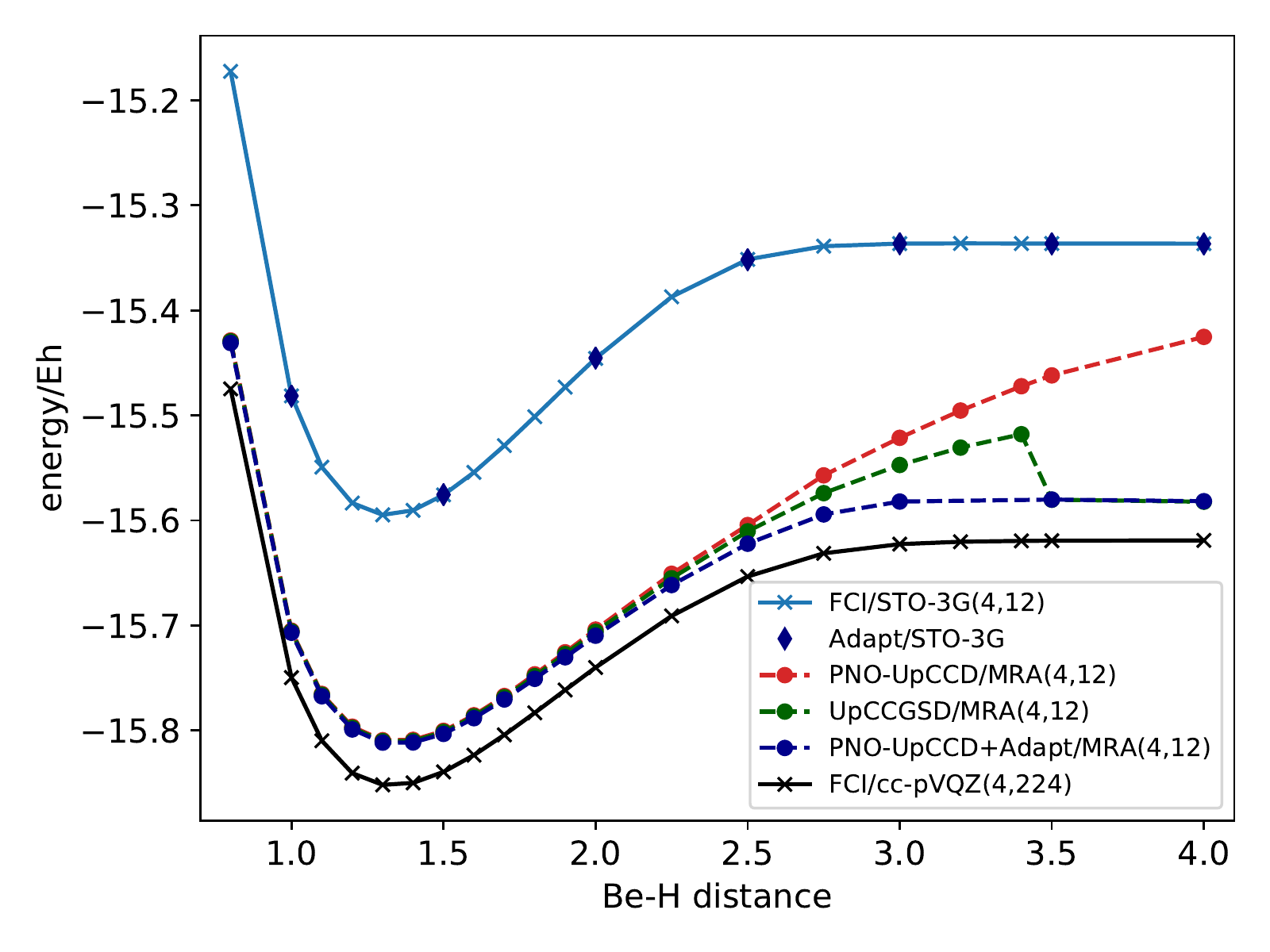}&
    \includegraphics[width=0.32\textwidth]{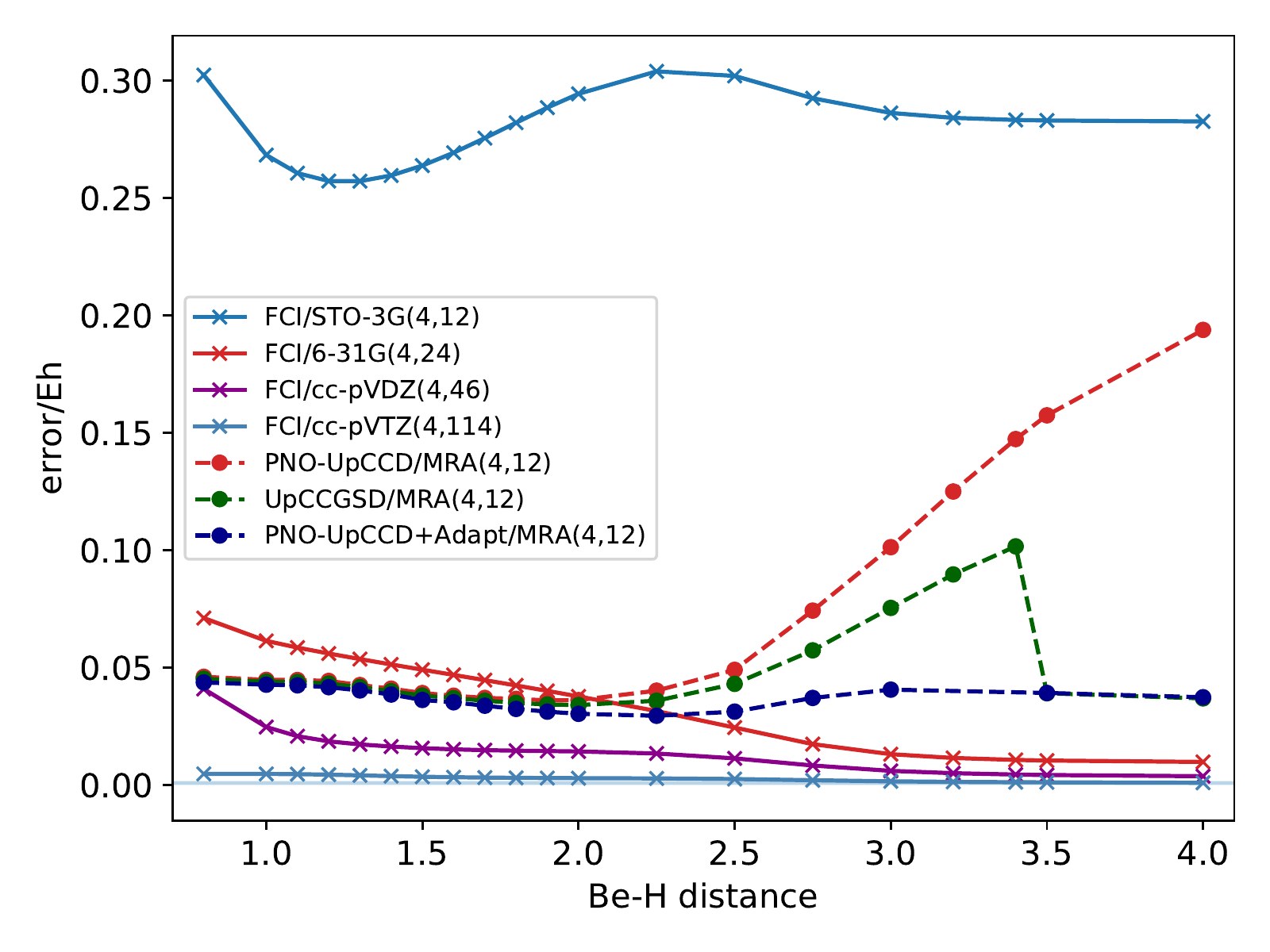}&
    \includegraphics[width=0.32\textwidth]{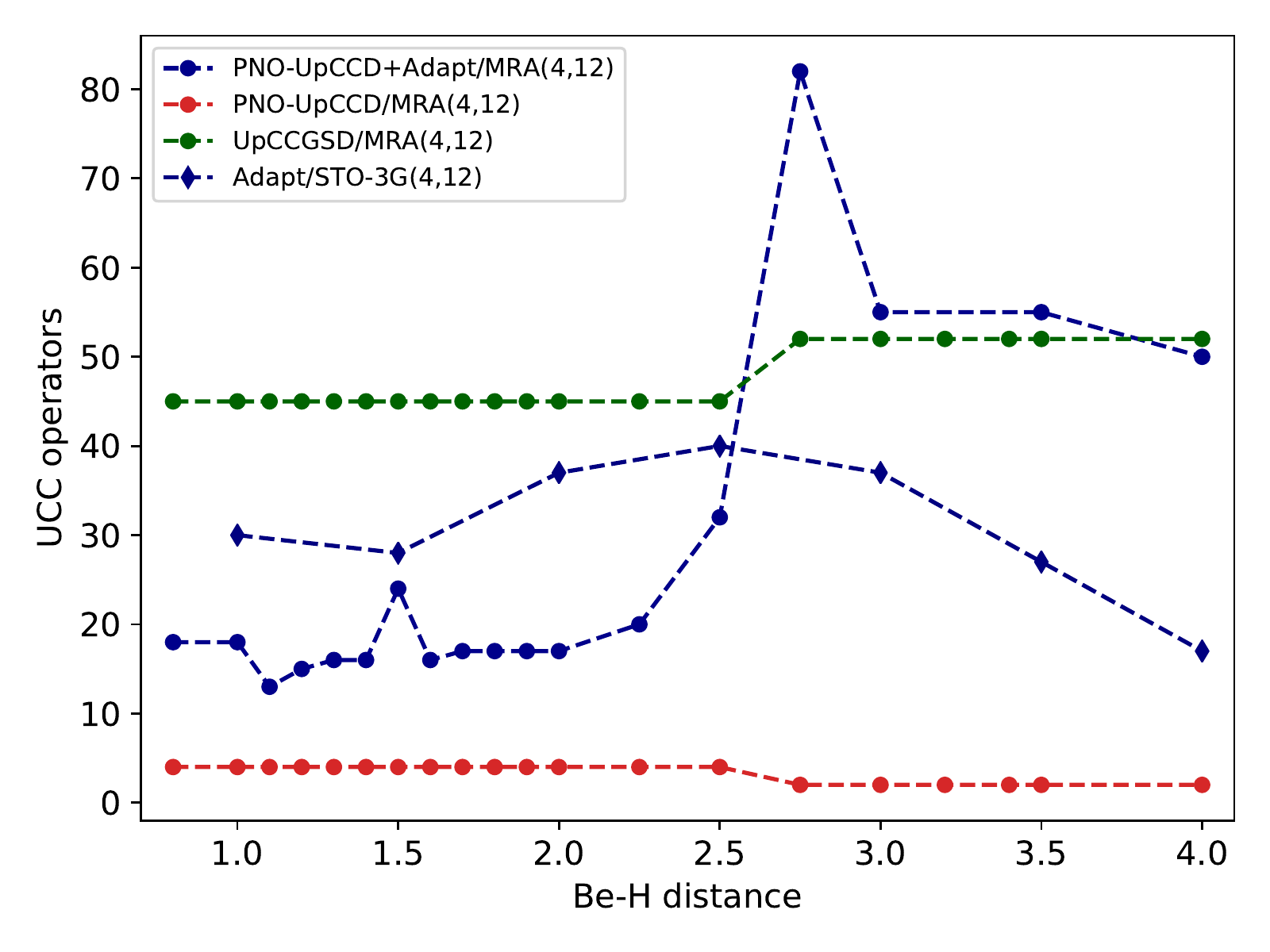}
    \end{tabular}
    \caption{\textbf{Bond dissociation curves} Total energies and errors with respect to the best affordable exact diagonalization (FCI) in the largest affordable Gaussian basis set (given in black). PNO-UpCC(S)D exploits the PNO structure of the surrogate model according to Eq.~\eqref{eq:def_pno-UpCCD}.  }
    \label{fig:pes_plots}
\end{figure*}

We simulated VQE energies, employing UpCCGSD, and the pair-specific PNO-UpCCD model, along the potential energy surfaces of the small molecules H$_2$, LiH, BH and BeH$_2$ and compared it with the best affordable variational methods using the large Gaussian basis sets. The potential energy surfaces are shown in Figs.~\ref{fig:h2_plot} and~\ref{fig:pes_plots}.
Similar as for the Helium atom in Fig.~\ref{fig:atoms} we show the PNO-MP2/MRA results in Fig.~\ref{fig:h2_plot} to illustrate that the problem is not fully solved by the surrogate model which performs poorly for the stretched geometries and the improvements in the method are in this case within the same range as the improvements in numerical representation.
For H$_2$ we furthermore show GBS calculations where the basis set was optimized for each point individually according to Ref.~\citenum{tamayo2018} using the \textsc{diffiqult}~\cite{diffiqult} package. 
This can be seen as an intermediate approach using Hartree{\textendash}Fock as a surrogate model and allowing the basis set to relax. 
In the (2,4) representation the optimal spatial orbitals can be well approximated by two atom-centered s-type orbitals, leading to significant improvements in the optimized STO-3G representation if individual optimization is enabled. 
For the slightly larger 6-31G basis set, individual optimization does not lead to improvements. 
In this case, the optimal (4,8) spatial orbitals contain two $\pi$ orbitals which intrinsically can not be represented by the 6-31G basis set but are well approximated by MRA where the techniques of Refs.~\citenum{kottmann2015numerically,kottmann2020direct} ensure that the correct symmetries are present in the initial guess functions for the MRA-PNOs.\\

In order to accurately describe chemistry, obtaining consistent relative energies over different molecular structures is in most cases more important than accurate absolute energies. In Fig.~\ref{fig:pes_npe_max}, we use non-parallelity (NPE) and maximum (MAX) errors as accuracy metrics with the best achievable method and Gaussian basis set as reference.
We chose the three diatomic molecules here such that UpCCGSD is a good ansatz, \textit{i.e.} differences to FCI are below the millihartree threshold for STO-3G and 6-31G simulations. {For the BeH$_2$ molecule, UpCCGSD did not always converge towards the best solution. We employed adaptive operator growth according to Refs.~\citenum{grimsley2019adaptive}~and~ \citenum{kottmann2020feasible}, in order to reach the lowest energies of the given PNO Hamiltonians.} As references, we used FCI/cc-pVQZ(2,120) for H$_2$,  FCI/cc-pCVQZ(2,288) for LiH, CISDTQ/cc-pCVTZ(6,114) for BH and FCI/cc-pVQZ(4,224) for BeH$_2$. Comparison with accurate numerical results from Refs.~\citenum{bande_lih_2010} and~\citenum{hiroshi2012} confirms, that the reference values for H$_2$ and LiH are close to the basis set limit.
In terms of NPE and MAX errors (Fig.~\ref{fig:pes_npe_max}), the basis-set-free VQE clearly outperforms the traditional basis sets with the same number of qubits as well as the cc-pVDZ simulations, which use approximately twice the number of qubits.
{The error of the MRA-PNO representation tends to grow with larger bond distances. This result is not unexpected, as the MP2 surrogate model performs worst in this regime. The effect on the orbitals determined by the surrogate model are however not that severe.
In contrast, the GBS representation performs best at large bond distances. This is also not surprising, since these basis sets are optimized for the atomic systems. The PNO-restricted unitaries perform well at not-too-far-stretched bond distances, where they offer significant savings in the number of operators used for the UCC ansatz (see Tab.~\ref{tab:parameter_counts}). This ansatz can be viewed as an even further restricted form of the approach in Refs.~\citenum{elfving2020simulating} and~\citenum{sokolov2020quantum}, where all configurations in the UCC wavefunction are restricted to double occupancies. In contrast to Ref.~\citenum{elfving2020simulating}, the PNO-UpCCD ansatz however performs well for all bond distances of LiH. The reason for this lies in the PNO structure, that is not present in canonical Hartree{\textendash}Fock orbitals. Note, that this is independent of the underlying numerical representation of the orbitals.}

\subsection{Umbrella reaction of Ammonia}
\begin{figure}
    \includegraphics[width=0.45\textwidth]{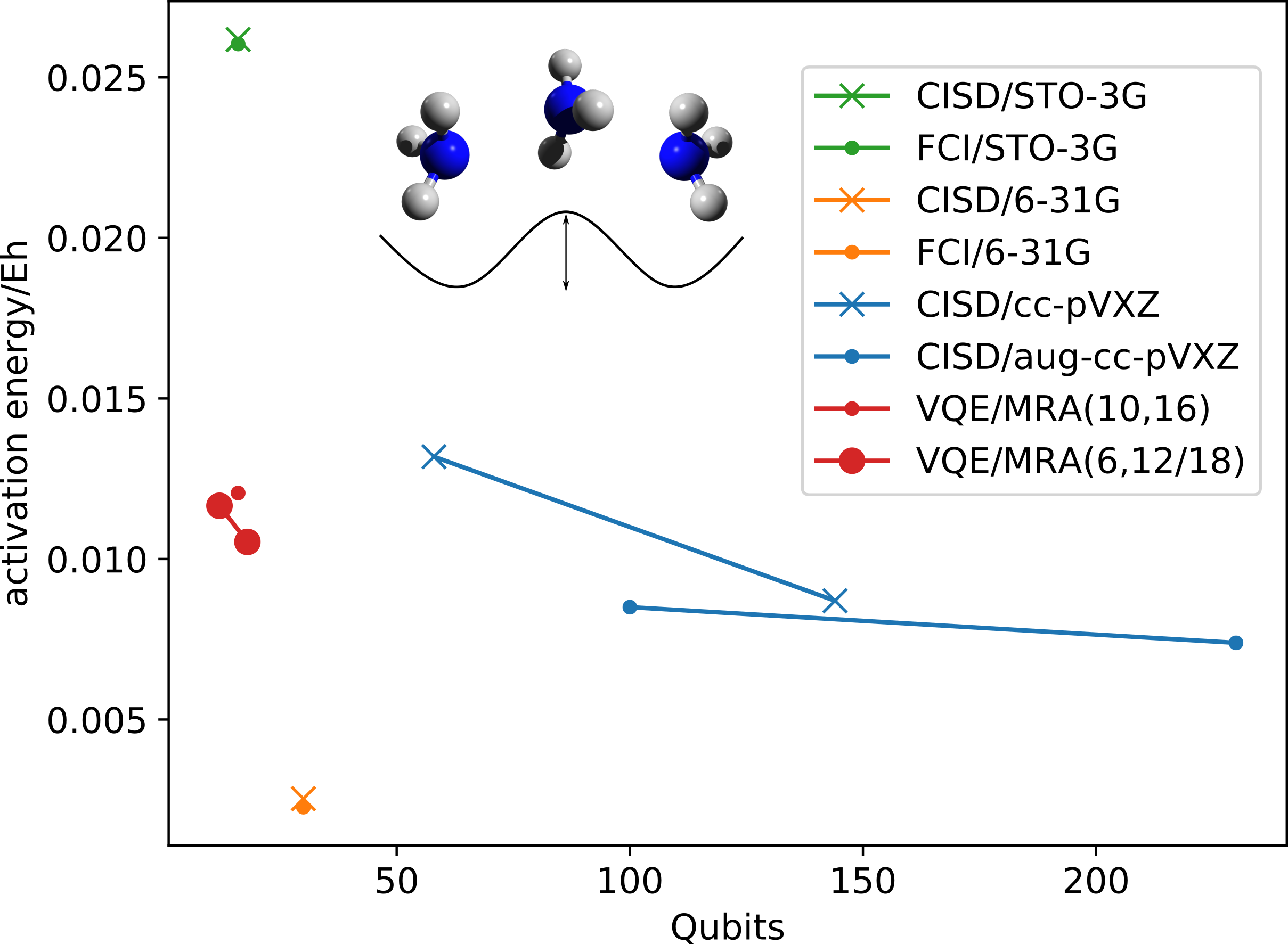}
    \caption{\textbf{Umbrella reaction of ammonia}: Energy barriers in relation to qubit requirements. Results are computed with different classical methods and the basis-set-free VQE (UpCCGSD ansatz).} 
    \label{fig:nh3_barrier}
\end{figure}
As a last example, we simulate the umbrella reaction of ammonia, a small intra-molecular reaction, where the umbrella-like molecular structure of ammonia is inverted passing a planar transition state.
The activation barriers of this reaction were simulated using the basis-set-free VQE approach and different classical methods with large basis sets, \textit{cf.} Fig.~\ref{fig:nh3_barrier}.
For larger basis sets, exact diagonalization is already unfeasible here. {CISD, however, provides an accurate model for this reaction type since possible size-inconsistency issues are negligible in this intramolecular reaction , wittnessed by the accompanied FCI calculations performed for the smaller basis-sets.}
Additional to the full treatment of the 10 electrons of ammonia, we used a 6 electron active space, freezing the lowest occupied Hartree{\textendash}Fock orbital and the orbital corresponding to the lone pair of ammonia.
Simulations with the small basis sets STO-3G and 6-31G over- or underestimate the activation barrier by more than a factor of two, while the basis-set-free approaches yields accurate energies. In this case, 12-18 qubits suffice for the basis-set-free approach to achieve a numerical accuracy, that would require 50-100 spin-orbitals with traditional basis sets. \\

\section{Conclusion \& Outlook}
In this work, we developed a basis-set-free approach by using MRA-PNO-MP2 as a surrogate model to construct compact, system-adapted qubit Hamiltonians with high numerical precision.
Initial demonstrations for absolute energies of atoms, small potential energy surfaces, and a toy model for a chemical reaction show a clear advantage of this approach compared to standard basis sets throughout all used metrics, allowing to achieve high numerical precision in the spatial representation of the wavefunction with significantly reduced number of qubits (see Tab.~\ref{tab:qubit_resources}).
Our approach is furthermore a promising candidate towards black-box quantum chemistry on quantum computers. In combination with adaptive circuit construction~\cite{ryabinkin2018qubit, grimsley2019adaptive}, this approach opens a path towards fully adaptive quantum chemistry independent of static basis sets and ansatz models. Combinations of adaptive circuit construction with generalized pair approximations~\cite{lee2018generalized, sokolov2020quantum} in the spirit of Ref.~\cite{kottmann2020feasible} are promising candidates towards accurate, adaptive, and computational feasible variational methods. 
Our approach is different in spirit as other approaches aiming to reduce the qubit requirements by applying external corrections~\cite{takeshita_increasing_2020} or by using symmetries to obtain compressed representations~\cite{bravyi2017tapering, setia2019reducing}. These approaches could be applied within the basis-set-free representation in the same way.
Further improvements on qubit requirements and numerical accuracy can be expected in the context of explicitly correlated approaches, that were already applied in the original formulation of the MRA-PNO-MP2 optimization.~\cite{kottmann2020direct} In the VQE context, approaches using trans-correlated Hamiltonians have been demonstrated recently using LCAO representations. These methods are promising candidates to further improve the performance of the basis-set-free VQE.
{We developed the first applications that exploit the structure of the pair-natural orbitals. Independent on the underlying numerical representation, this leads to low-depth quantum circuits, allowing savings in the qubit and gate requirements. These approaches have the potential to play an important role in future developments, either as benchmarks for small quantum hardware, or as initial states for more sophisticated quantum algorithms as illustrated for BeH$_2$ in Fig.~\ref{fig:pes_plots}. }
In this work, we used MP2 as correlated surrogate model, but other models are also possible within this framework, and might bring additional advantages. A natural extension to MP2 is given by the coupled-cluster hierarchy, where models like CCD and CCSD first come to mind. More restricted models like pair-restricted coupled-cluster, that are also employed in quantum circuit construction~\cite{sokolov2020quantum, lee2018generalized}, offer advantages through lower computational cost. In principle, more advanced surrogate models, that for example incorporate higher-order coupled-cluster terms only for the determination of specific orbitals, can be envisioned as well.\\

\textbf{Notes} Our implementations are open-source and available online through the \textsc{tequila} package under \href{https://github.com/aspuru-guzik-group/tequila}{github.com/aspuru-guzik-group/tequila}, where we provide initial tutorials on the usage. A modified version of the MRA-PNO-MP2 implementation is available over a separate \textsc{madness} branch under \href{https://github.com/kottmanj/madness/tree/tequila}{github.com/kottmanj/madness/tree/tequila}. Feel free to contact JSK for more information and updates. The authors declare no competing financial interest. 

\section{Acknowledgement}
This work was supported by the U.S. Department of Energy under Award No. DE-SC0019374.
A.A.-G. acknowledges the generous support from Google, Inc.  in the form of a Google Focused Award. A.A.-G. also acknowledges support from the Canada Industrial Research Chairs  Program and the Canada 150 Research Chairs Program. We thank the generous support of Anders G. Fr\o{}seth.
P.S. acknowledges support by a fellowship within the IFI programme of the German Academic Exchange Service (DAAD). This research was enabled in part by support provided by Compute Canada. Computations were performed on the niagara supercomputer at the SciNet HPC Consortium.~\cite{niagara1, niagara2} SciNet is funded by: the Canada Foundation for Innovation; the Government of Ontario; Ontario Research Fund - Research Excellence; and the University of Toronto.
\bibliography{main.bib}

\begin{thebibliography}{78}%
\makeatletter
\providecommand \@ifxundefined [1]{%
 \@ifx{#1\undefined}
}%
\providecommand \@ifnum [1]{%
 \ifnum #1\expandafter \@firstoftwo
 \else \expandafter \@secondoftwo
 \fi
}%
\providecommand \@ifx [1]{%
 \ifx #1\expandafter \@firstoftwo
 \else \expandafter \@secondoftwo
 \fi
}%
\providecommand \natexlab [1]{#1}%
\providecommand \enquote  [1]{``#1''}%
\providecommand \bibnamefont  [1]{#1}%
\providecommand \bibfnamefont [1]{#1}%
\providecommand \citenamefont [1]{#1}%
\providecommand \href@noop [0]{\@secondoftwo}%
\providecommand \href [0]{\begingroup \@sanitize@url \@href}%
\providecommand \@href[1]{\@@startlink{#1}\@@href}%
\providecommand \@@href[1]{\endgroup#1\@@endlink}%
\providecommand \@sanitize@url [0]{\catcode `\\12\catcode `\$12\catcode
  `\&12\catcode `\#12\catcode `\^12\catcode `\_12\catcode `\%12\relax}%
\providecommand \@@startlink[1]{}%
\providecommand \@@endlink[0]{}%
\providecommand \url  [0]{\begingroup\@sanitize@url \@url }%
\providecommand \@url [1]{\endgroup\@href {#1}{\urlprefix }}%
\providecommand \urlprefix  [0]{URL }%
\providecommand \Eprint [0]{\href }%
\providecommand \doibase [0]{https://doi.org/}%
\providecommand \selectlanguage [0]{\@gobble}%
\providecommand \bibinfo  [0]{\@secondoftwo}%
\providecommand \bibfield  [0]{\@secondoftwo}%
\providecommand \translation [1]{[#1]}%
\providecommand \BibitemOpen [0]{}%
\providecommand \bibitemStop [0]{}%
\providecommand \bibitemNoStop [0]{.\EOS\space}%
\providecommand \EOS [0]{\spacefactor3000\relax}%
\providecommand \BibitemShut  [1]{\csname bibitem#1\endcsname}%
\let\auto@bib@innerbib\@empty
\bibitem [{\citenamefont {Aspuru-Guzik}\ \emph {et~al.}(2018)\citenamefont
  {Aspuru-Guzik}, \citenamefont {Lindh},\ and\ \citenamefont
  {Reiher}}]{aag2018revolution}%
  \BibitemOpen
  \bibfield  {author} {\bibinfo {author} {\bibfnamefont {A.}~\bibnamefont
  {Aspuru-Guzik}}, \bibinfo {author} {\bibfnamefont {R.}~\bibnamefont
  {Lindh}},\ and\ \bibinfo {author} {\bibfnamefont {M.}~\bibnamefont
  {Reiher}},\ }\bibfield  {title} {\bibinfo {title} {The {M}atter {S}imulation
  {(R)}evolution},\ }\href {https://doi.org/10.1021/acscentsci.7b00550}
  {\bibfield  {journal} {\bibinfo  {journal} {ACS Central Science}\ }\textbf
  {\bibinfo {volume} {4}},\ \bibinfo {pages} {144} (\bibinfo {year} {2018})},\
  \bibinfo {note} {pMID: 29532014},\ \Eprint
  {https://arxiv.org/abs/https://doi.org/10.1021/acscentsci.7b00550}
  {https://doi.org/10.1021/acscentsci.7b00550} \BibitemShut {NoStop}%
\bibitem [{\citenamefont {Fermann}\ and\ \citenamefont
  {Valeev}(2020)}]{fermann2020fundamentals}%
  \BibitemOpen
  \bibfield  {author} {\bibinfo {author} {\bibfnamefont {J.~T.}\ \bibnamefont
  {Fermann}}\ and\ \bibinfo {author} {\bibfnamefont {E.~F.}\ \bibnamefont
  {Valeev}},\ }\href@noop {} {\bibinfo {title} {Fundamentals of molecular
  integrals evaluation}} (\bibinfo {year} {2020}),\ \Eprint
  {https://arxiv.org/abs/2007.12057} {arXiv:2007.12057 [quant-ph]} \BibitemShut
  {NoStop}%
\bibitem [{\citenamefont {Helgaker}\ \emph {et~al.}(2014)\citenamefont
  {Helgaker}, \citenamefont {Jorgensen},\ and\ \citenamefont
  {Olsen}}]{helgaker2014molecular}%
  \BibitemOpen
  \bibfield  {author} {\bibinfo {author} {\bibfnamefont {T.}~\bibnamefont
  {Helgaker}}, \bibinfo {author} {\bibfnamefont {P.}~\bibnamefont
  {Jorgensen}},\ and\ \bibinfo {author} {\bibfnamefont {J.}~\bibnamefont
  {Olsen}},\ }\href@noop {} {\emph {\bibinfo {title} {Molecular
  electronic-structure theory}}}\ (\bibinfo  {publisher} {John Wiley \& Sons},\
  \bibinfo {year} {2014})\BibitemShut {NoStop}%
\bibitem [{\citenamefont {Pritchard}\ \emph {et~al.}(2019)\citenamefont
  {Pritchard}, \citenamefont {Altarawy}, \citenamefont {Didier}, \citenamefont
  {Gibson},\ and\ \citenamefont {Windus}}]{pritchard2019new}%
  \BibitemOpen
  \bibfield  {author} {\bibinfo {author} {\bibfnamefont {B.~P.}\ \bibnamefont
  {Pritchard}}, \bibinfo {author} {\bibfnamefont {D.}~\bibnamefont {Altarawy}},
  \bibinfo {author} {\bibfnamefont {B.}~\bibnamefont {Didier}}, \bibinfo
  {author} {\bibfnamefont {T.~D.}\ \bibnamefont {Gibson}},\ and\ \bibinfo
  {author} {\bibfnamefont {T.~L.}\ \bibnamefont {Windus}},\ }\bibfield  {title}
  {\bibinfo {title} {New basis set exchange: An open, up-to-date resource for
  the molecular sciences community},\ }\href
  {https://doi.org/10.1021/acs.jcim.9b00725} {\bibfield  {journal} {\bibinfo
  {journal} {Journal of chemical information and modeling}\ }\textbf {\bibinfo
  {volume} {59}},\ \bibinfo {pages} {4814} (\bibinfo {year}
  {2019})}\BibitemShut {NoStop}%
\bibitem [{\citenamefont {Van~Lenthe}\ and\ \citenamefont
  {Baerends}(2003)}]{lenthe2003}%
  \BibitemOpen
  \bibfield  {author} {\bibinfo {author} {\bibfnamefont {E.}~\bibnamefont
  {Van~Lenthe}}\ and\ \bibinfo {author} {\bibfnamefont {E.~J.}\ \bibnamefont
  {Baerends}},\ }\bibfield  {title} {\bibinfo {title} {Optimized slater-type
  basis sets for the elements 1–118},\ }\href
  {https://doi.org/10.1002/jcc.10255} {\bibfield  {journal} {\bibinfo
  {journal} {Journal of Computational Chemistry}\ }\textbf {\bibinfo {volume}
  {24}},\ \bibinfo {pages} {1142} (\bibinfo {year} {2003})},\ \Eprint
  {https://arxiv.org/abs/https://onlinelibrary.wiley.com/doi/pdf/10.1002/jcc.10255}
  {https://onlinelibrary.wiley.com/doi/pdf/10.1002/jcc.10255} \BibitemShut
  {NoStop}%
\bibitem [{\citenamefont {Herbst}\ \emph {et~al.}(2019)\citenamefont {Herbst},
  \citenamefont {Avery},\ and\ \citenamefont {Dreuw}}]{herbst2019}%
  \BibitemOpen
  \bibfield  {author} {\bibinfo {author} {\bibfnamefont {M.~F.}\ \bibnamefont
  {Herbst}}, \bibinfo {author} {\bibfnamefont {J.~E.}\ \bibnamefont {Avery}},\
  and\ \bibinfo {author} {\bibfnamefont {A.}~\bibnamefont {Dreuw}},\ }\bibfield
   {title} {\bibinfo {title} {Quantum chemistry with coulomb sturmians:
  Construction and convergence of coulomb sturmian basis sets at the
  hartree-fock level},\ }\href {https://doi.org/10.1103/PhysRevA.99.012512}
  {\bibfield  {journal} {\bibinfo  {journal} {Phys. Rev. A}\ }\textbf {\bibinfo
  {volume} {99}},\ \bibinfo {pages} {012512} (\bibinfo {year}
  {2019})}\BibitemShut {NoStop}%
\bibitem [{\citenamefont {Genovese}\ \emph {et~al.}(2008)\citenamefont
  {Genovese}, \citenamefont {Neelov}, \citenamefont {Goedecker}, \citenamefont
  {Deutsch}, \citenamefont {Ghasemi}, \citenamefont {Willand}, \citenamefont
  {Caliste}, \citenamefont {Zilberberg}, \citenamefont {Rayson}, \citenamefont
  {Bergman},\ and\ \citenamefont {Schneider}}]{genovese2008bigdft}%
  \BibitemOpen
  \bibfield  {author} {\bibinfo {author} {\bibfnamefont {L.}~\bibnamefont
  {Genovese}}, \bibinfo {author} {\bibfnamefont {A.}~\bibnamefont {Neelov}},
  \bibinfo {author} {\bibfnamefont {S.}~\bibnamefont {Goedecker}}, \bibinfo
  {author} {\bibfnamefont {T.}~\bibnamefont {Deutsch}}, \bibinfo {author}
  {\bibfnamefont {S.~A.}\ \bibnamefont {Ghasemi}}, \bibinfo {author}
  {\bibfnamefont {A.}~\bibnamefont {Willand}}, \bibinfo {author} {\bibfnamefont
  {D.}~\bibnamefont {Caliste}}, \bibinfo {author} {\bibfnamefont
  {O.}~\bibnamefont {Zilberberg}}, \bibinfo {author} {\bibfnamefont
  {M.}~\bibnamefont {Rayson}}, \bibinfo {author} {\bibfnamefont
  {A.}~\bibnamefont {Bergman}},\ and\ \bibinfo {author} {\bibfnamefont
  {R.}~\bibnamefont {Schneider}},\ }\bibfield  {title} {\bibinfo {title}
  {Daubechies wavelets as a basis set for density functional pseudopotential
  calculations},\ }\href {https://doi.org/10.1063/1.2949547} {\bibfield
  {journal} {\bibinfo  {journal} {The Journal of Chemical Physics}\ }\textbf
  {\bibinfo {volume} {129}},\ \bibinfo {pages} {014109} (\bibinfo {year}
  {2008})},\ \Eprint {https://arxiv.org/abs/https://doi.org/10.1063/1.2949547}
  {https://doi.org/10.1063/1.2949547} \BibitemShut {NoStop}%
\bibitem [{\citenamefont {Mohr}\ \emph {et~al.}(2015)\citenamefont {Mohr},
  \citenamefont {Ratcliff}, \citenamefont {Genovese}, \citenamefont {Caliste},
  \citenamefont {Boulanger}, \citenamefont {Goedecker},\ and\ \citenamefont
  {Deutsch}}]{mohr2015bigdft}%
  \BibitemOpen
  \bibfield  {author} {\bibinfo {author} {\bibfnamefont {S.}~\bibnamefont
  {Mohr}}, \bibinfo {author} {\bibfnamefont {L.~E.}\ \bibnamefont {Ratcliff}},
  \bibinfo {author} {\bibfnamefont {L.}~\bibnamefont {Genovese}}, \bibinfo
  {author} {\bibfnamefont {D.}~\bibnamefont {Caliste}}, \bibinfo {author}
  {\bibfnamefont {P.}~\bibnamefont {Boulanger}}, \bibinfo {author}
  {\bibfnamefont {S.}~\bibnamefont {Goedecker}},\ and\ \bibinfo {author}
  {\bibfnamefont {T.}~\bibnamefont {Deutsch}},\ }\bibfield  {title} {\bibinfo
  {title} {Accurate and efficient linear scaling dft calculations with
  universal applicability},\ }\href {https://doi.org/10.1039/C5CP00437C}
  {\bibfield  {journal} {\bibinfo  {journal} {Phys. Chem. Chem. Phys.}\
  }\textbf {\bibinfo {volume} {17}},\ \bibinfo {pages} {31360} (\bibinfo {year}
  {2015})}\BibitemShut {NoStop}%
\bibitem [{\citenamefont {Ratcliff}\ \emph {et~al.}(2020)\citenamefont
  {Ratcliff}, \citenamefont {Dawson}, \citenamefont {Fisicaro}, \citenamefont
  {Caliste}, \citenamefont {Mohr}, \citenamefont {Degomme}, \citenamefont
  {Videau}, \citenamefont {Cristiglio}, \citenamefont {Stella}, \citenamefont
  {D’Alessandro}, \citenamefont {Goedecker}, \citenamefont {Nakajima},
  \citenamefont {Deutsch},\ and\ \citenamefont {Genovese}}]{ratcliff2020}%
  \BibitemOpen
  \bibfield  {author} {\bibinfo {author} {\bibfnamefont {L.~E.}\ \bibnamefont
  {Ratcliff}}, \bibinfo {author} {\bibfnamefont {W.}~\bibnamefont {Dawson}},
  \bibinfo {author} {\bibfnamefont {G.}~\bibnamefont {Fisicaro}}, \bibinfo
  {author} {\bibfnamefont {D.}~\bibnamefont {Caliste}}, \bibinfo {author}
  {\bibfnamefont {S.}~\bibnamefont {Mohr}}, \bibinfo {author} {\bibfnamefont
  {A.}~\bibnamefont {Degomme}}, \bibinfo {author} {\bibfnamefont
  {B.}~\bibnamefont {Videau}}, \bibinfo {author} {\bibfnamefont
  {V.}~\bibnamefont {Cristiglio}}, \bibinfo {author} {\bibfnamefont
  {M.}~\bibnamefont {Stella}}, \bibinfo {author} {\bibfnamefont
  {M.}~\bibnamefont {D’Alessandro}}, \bibinfo {author} {\bibfnamefont
  {S.}~\bibnamefont {Goedecker}}, \bibinfo {author} {\bibfnamefont
  {T.}~\bibnamefont {Nakajima}}, \bibinfo {author} {\bibfnamefont
  {T.}~\bibnamefont {Deutsch}},\ and\ \bibinfo {author} {\bibfnamefont
  {L.}~\bibnamefont {Genovese}},\ }\bibfield  {title} {\bibinfo {title}
  {Flexibilities of wavelets as a computational basis set for large-scale
  electronic structure calculations},\ }\href
  {https://doi.org/10.1063/5.0004792} {\bibfield  {journal} {\bibinfo
  {journal} {The Journal of Chemical Physics}\ }\textbf {\bibinfo {volume}
  {152}},\ \bibinfo {pages} {194110} (\bibinfo {year} {2020})},\ \Eprint
  {https://arxiv.org/abs/https://doi.org/10.1063/5.0004792}
  {https://doi.org/10.1063/5.0004792} \BibitemShut {NoStop}%
\bibitem [{\citenamefont {Kang}\ \emph {et~al.}(2020)\citenamefont {Kang},
  \citenamefont {Woo}, \citenamefont {Kim}, \citenamefont {Kim}, \citenamefont
  {Kim}, \citenamefont {Lim}, \citenamefont {Choi},\ and\ \citenamefont
  {Kim}}]{Kang2020}%
  \BibitemOpen
  \bibfield  {author} {\bibinfo {author} {\bibfnamefont {S.}~\bibnamefont
  {Kang}}, \bibinfo {author} {\bibfnamefont {J.}~\bibnamefont {Woo}}, \bibinfo
  {author} {\bibfnamefont {J.}~\bibnamefont {Kim}}, \bibinfo {author}
  {\bibfnamefont {H.}~\bibnamefont {Kim}}, \bibinfo {author} {\bibfnamefont
  {Y.}~\bibnamefont {Kim}}, \bibinfo {author} {\bibfnamefont {J.}~\bibnamefont
  {Lim}}, \bibinfo {author} {\bibfnamefont {S.}~\bibnamefont {Choi}},\ and\
  \bibinfo {author} {\bibfnamefont {W.~Y.}\ \bibnamefont {Kim}},\ }\bibfield
  {title} {\bibinfo {title} {Ace-molecule: An open-source real-space quantum
  chemistry package},\ }\href {https://doi.org/10.1063/5.0002959} {\bibfield
  {journal} {\bibinfo  {journal} {The Journal of Chemical Physics}\ }\textbf
  {\bibinfo {volume} {152}},\ \bibinfo {pages} {124110} (\bibinfo {year}
  {2020})},\ \Eprint {https://arxiv.org/abs/https://doi.org/10.1063/5.0002959}
  {https://doi.org/10.1063/5.0002959} \BibitemShut {NoStop}%
\bibitem [{\citenamefont {Harrison}\ \emph {et~al.}(2004)\citenamefont
  {Harrison}, \citenamefont {Fann}, \citenamefont {Yanai}, \citenamefont
  {Gan},\ and\ \citenamefont {Beylkin}}]{harrison2004multiresolution}%
  \BibitemOpen
  \bibfield  {author} {\bibinfo {author} {\bibfnamefont {R.~J.}\ \bibnamefont
  {Harrison}}, \bibinfo {author} {\bibfnamefont {G.~I.}\ \bibnamefont {Fann}},
  \bibinfo {author} {\bibfnamefont {T.}~\bibnamefont {Yanai}}, \bibinfo
  {author} {\bibfnamefont {Z.}~\bibnamefont {Gan}},\ and\ \bibinfo {author}
  {\bibfnamefont {G.}~\bibnamefont {Beylkin}},\ }\bibfield  {title} {\bibinfo
  {title} {Multiresolution quantum chemistry: Basic theory and initial
  applications},\ }\href {https://doi.org/10.1063/1.1791051} {\bibfield
  {journal} {\bibinfo  {journal} {The Journal of chemical physics}\ }\textbf
  {\bibinfo {volume} {121}},\ \bibinfo {pages} {11587} (\bibinfo {year}
  {2004})}\BibitemShut {NoStop}%
\bibitem [{\citenamefont {Harrison}\ \emph {et~al.}(2016)\citenamefont
  {Harrison}, \citenamefont {Beylkin}, \citenamefont {Bischoff}, \citenamefont
  {Calvin}, \citenamefont {Fann}, \citenamefont {Fosso-Tande}, \citenamefont
  {Galindo}, \citenamefont {Hammond}, \citenamefont {Hartman-Baker},
  \citenamefont {Hill} \emph {et~al.}}]{harrison2016madness}%
  \BibitemOpen
  \bibfield  {author} {\bibinfo {author} {\bibfnamefont {R.~J.}\ \bibnamefont
  {Harrison}}, \bibinfo {author} {\bibfnamefont {G.}~\bibnamefont {Beylkin}},
  \bibinfo {author} {\bibfnamefont {F.~A.}\ \bibnamefont {Bischoff}}, \bibinfo
  {author} {\bibfnamefont {J.~A.}\ \bibnamefont {Calvin}}, \bibinfo {author}
  {\bibfnamefont {G.~I.}\ \bibnamefont {Fann}}, \bibinfo {author}
  {\bibfnamefont {J.}~\bibnamefont {Fosso-Tande}}, \bibinfo {author}
  {\bibfnamefont {D.}~\bibnamefont {Galindo}}, \bibinfo {author} {\bibfnamefont
  {J.~R.}\ \bibnamefont {Hammond}}, \bibinfo {author} {\bibfnamefont
  {R.}~\bibnamefont {Hartman-Baker}}, \bibinfo {author} {\bibfnamefont {J.~C.}\
  \bibnamefont {Hill}}, \emph {et~al.},\ }\bibfield  {title} {\bibinfo {title}
  {Madness: A multiresolution, adaptive numerical environment for scientific
  simulation},\ }\href {https://doi.org/10.1137/15M1026171} {\bibfield
  {journal} {\bibinfo  {journal} {SIAM Journal on Scientific Computing}\
  }\textbf {\bibinfo {volume} {38}},\ \bibinfo {pages} {S123} (\bibinfo {year}
  {2016})}\BibitemShut {NoStop}%
\bibitem [{\citenamefont {Bischoff}(2019)}]{bischoff2019computing}%
  \BibitemOpen
  \bibfield  {author} {\bibinfo {author} {\bibfnamefont {F.~A.}\ \bibnamefont
  {Bischoff}},\ }\bibfield  {title} {\bibinfo {title} {Computing accurate
  molecular properties in real space using multiresolution analysis},\ }in\
  \href
  {https://www.elsevier.com/books/state-of-the-art-of-molecular-electronic-structure-computations-correlation-methods-basis-sets-and-more/hoggan/978-0-12-816174-6}
  {\emph {\bibinfo {booktitle} {Advances in Quantum Chemistry}}},\
  Vol.~\bibinfo {volume} {79}\ (\bibinfo  {publisher} {Elsevier},\ \bibinfo
  {year} {2019})\ pp.\ \bibinfo {pages} {3--52}\BibitemShut {NoStop}%
\bibitem [{\citenamefont {Frediani}\ \emph {et~al.}(2013)\citenamefont
  {Frediani}, \citenamefont {Fossgaard}, \citenamefont {Flå},\ and\
  \citenamefont {Ruud}}]{Frediani2013}%
  \BibitemOpen
  \bibfield  {author} {\bibinfo {author} {\bibfnamefont {L.}~\bibnamefont
  {Frediani}}, \bibinfo {author} {\bibfnamefont {E.}~\bibnamefont {Fossgaard}},
  \bibinfo {author} {\bibfnamefont {T.}~\bibnamefont {Flå}},\ and\ \bibinfo
  {author} {\bibfnamefont {K.}~\bibnamefont {Ruud}},\ }\bibfield  {title}
  {\bibinfo {title} {Fully adaptive algorithms for multivariate integral
  equations using the non-standard form and multiwavelets with applications to
  the poisson and bound-state helmholtz kernels in three dimensions},\ }\href
  {https://doi.org/10.1080/00268976.2013.810793} {\bibfield  {journal}
  {\bibinfo  {journal} {Molecular Physics}\ }\textbf {\bibinfo {volume}
  {111}},\ \bibinfo {pages} {1143} (\bibinfo {year} {2013})},\ \Eprint
  {https://arxiv.org/abs/https://doi.org/10.1080/00268976.2013.810793}
  {https://doi.org/10.1080/00268976.2013.810793} \BibitemShut {NoStop}%
\bibitem [{\citenamefont {Jensen}\ \emph {et~al.}(2017)\citenamefont {Jensen},
  \citenamefont {Saha}, \citenamefont {Flores-Livas}, \citenamefont {Huhn},
  \citenamefont {Blum}, \citenamefont {Goedecker},\ and\ \citenamefont
  {Frediani}}]{jensen2017elephant}%
  \BibitemOpen
  \bibfield  {author} {\bibinfo {author} {\bibfnamefont {S.~R.}\ \bibnamefont
  {Jensen}}, \bibinfo {author} {\bibfnamefont {S.}~\bibnamefont {Saha}},
  \bibinfo {author} {\bibfnamefont {J.~A.}\ \bibnamefont {Flores-Livas}},
  \bibinfo {author} {\bibfnamefont {W.}~\bibnamefont {Huhn}}, \bibinfo {author}
  {\bibfnamefont {V.}~\bibnamefont {Blum}}, \bibinfo {author} {\bibfnamefont
  {S.}~\bibnamefont {Goedecker}},\ and\ \bibinfo {author} {\bibfnamefont
  {L.}~\bibnamefont {Frediani}},\ }\bibfield  {title} {\bibinfo {title} {The
  elephant in the room of density functional theory calculations},\ }\href
  {https://doi.org/10.1021/acs.jpclett.7b00255} {\bibfield  {journal} {\bibinfo
   {journal} {The Journal of Physical Chemistry Letters}\ }\textbf {\bibinfo
  {volume} {8}},\ \bibinfo {pages} {1449} (\bibinfo {year} {2017})},\ \bibinfo
  {note} {pMID: 28291362},\ \Eprint
  {https://arxiv.org/abs/https://doi.org/10.1021/acs.jpclett.7b00255}
  {https://doi.org/10.1021/acs.jpclett.7b00255} \BibitemShut {NoStop}%
\bibitem [{\citenamefont {Kottmann}\ \emph {et~al.}(2015)\citenamefont
  {Kottmann}, \citenamefont {H{\"o}fener},\ and\ \citenamefont
  {Bischoff}}]{kottmann2015numerically}%
  \BibitemOpen
  \bibfield  {author} {\bibinfo {author} {\bibfnamefont {J.~S.}\ \bibnamefont
  {Kottmann}}, \bibinfo {author} {\bibfnamefont {S.}~\bibnamefont
  {H{\"o}fener}},\ and\ \bibinfo {author} {\bibfnamefont {F.~A.}\ \bibnamefont
  {Bischoff}},\ }\bibfield  {title} {\bibinfo {title} {Numerically accurate
  linear response-properties in the configuration-interaction singles (cis)
  approximation},\ }\href {https://doi.org/10.1039/C5CP00345H} {\bibfield
  {journal} {\bibinfo  {journal} {Physical Chemistry Chemical Physics}\
  }\textbf {\bibinfo {volume} {17}},\ \bibinfo {pages} {31453} (\bibinfo {year}
  {2015})}\BibitemShut {NoStop}%
\bibitem [{\citenamefont {Yanai}\ \emph {et~al.}(2015)\citenamefont {Yanai},
  \citenamefont {Fann}, \citenamefont {Beylkin},\ and\ \citenamefont
  {Harrison}}]{yanai2015multiresolution}%
  \BibitemOpen
  \bibfield  {author} {\bibinfo {author} {\bibfnamefont {T.}~\bibnamefont
  {Yanai}}, \bibinfo {author} {\bibfnamefont {G.~I.}\ \bibnamefont {Fann}},
  \bibinfo {author} {\bibfnamefont {G.}~\bibnamefont {Beylkin}},\ and\ \bibinfo
  {author} {\bibfnamefont {R.~J.}\ \bibnamefont {Harrison}},\ }\bibfield
  {title} {\bibinfo {title} {Multiresolution quantum chemistry in multiwavelet
  bases: excited states from time-dependent hartree--fock and density
  functional theory via linear response},\ }\href
  {https://doi.org/10.1039/C4CP05821F} {\bibfield  {journal} {\bibinfo
  {journal} {Physical Chemistry Chemical Physics}\ }\textbf {\bibinfo {volume}
  {17}},\ \bibinfo {pages} {31405} (\bibinfo {year} {2015})}\BibitemShut
  {NoStop}%
\bibitem [{\citenamefont {Brakestad}\ \emph {et~al.}(2020)\citenamefont
  {Brakestad}, \citenamefont {Jensen}, \citenamefont {Wind}, \citenamefont
  {D’Alessandro}, \citenamefont {Genovese}, \citenamefont {Hopmann},\ and\
  \citenamefont {Frediani}}]{brakestad2020}%
  \BibitemOpen
  \bibfield  {author} {\bibinfo {author} {\bibfnamefont {A.}~\bibnamefont
  {Brakestad}}, \bibinfo {author} {\bibfnamefont {S.~R.}\ \bibnamefont
  {Jensen}}, \bibinfo {author} {\bibfnamefont {P.}~\bibnamefont {Wind}},
  \bibinfo {author} {\bibfnamefont {M.}~\bibnamefont {D’Alessandro}},
  \bibinfo {author} {\bibfnamefont {L.}~\bibnamefont {Genovese}}, \bibinfo
  {author} {\bibfnamefont {K.~H.}\ \bibnamefont {Hopmann}},\ and\ \bibinfo
  {author} {\bibfnamefont {L.}~\bibnamefont {Frediani}},\ }\bibfield  {title}
  {\bibinfo {title} {Static polarizabilities at the basis set limit: A
  benchmark of 124 species},\ }\href {https://doi.org/10.1021/acs.jctc.0c00128}
  {\bibfield  {journal} {\bibinfo  {journal} {Journal of Chemical Theory and
  Computation}\ }\textbf {\bibinfo {volume} {0}},\ \bibinfo {pages} {null}
  (\bibinfo {year} {2020})},\ \bibinfo {note} {pMID: 32544327},\ \Eprint
  {https://arxiv.org/abs/https://doi.org/10.1021/acs.jctc.0c00128}
  {https://doi.org/10.1021/acs.jctc.0c00128} \BibitemShut {NoStop}%
\bibitem [{\citenamefont {Sekino}\ \emph {et~al.}(2012)\citenamefont {Sekino},
  \citenamefont {Yokoi},\ and\ \citenamefont {Harrison}}]{sekino2012new}%
  \BibitemOpen
  \bibfield  {author} {\bibinfo {author} {\bibfnamefont {H.}~\bibnamefont
  {Sekino}}, \bibinfo {author} {\bibfnamefont {Y.}~\bibnamefont {Yokoi}},\ and\
  \bibinfo {author} {\bibfnamefont {R.~J.}\ \bibnamefont {Harrison}},\
  }\bibfield  {title} {\bibinfo {title} {A new implementation of dynamic
  polarizability evaluation using a multi-resolution multi-wavelet basis set},\
  }in\ \href {https://doi.org/10.1088%2F1742-6596%2F352%2F1%2F012014} {\emph
  {\bibinfo {booktitle} {Journal of Physics. Conference Series (Online)}}},\
  Vol.\ \bibinfo {volume} {352}\ (\bibinfo {year} {2012})\BibitemShut {NoStop}%
\bibitem [{\citenamefont {Bischoff}(2020)}]{bischoff2020magnetic}%
  \BibitemOpen
  \bibfield  {author} {\bibinfo {author} {\bibfnamefont {F.~A.}\ \bibnamefont
  {Bischoff}},\ }\bibfield  {title} {\bibinfo {title} {Structure of the
  ${\mathrm{h}}_{3}$ molecule in a strong homogeneous magnetic field as
  computed by the hartree-fock method using multiresolution analysis},\ }\href
  {https://doi.org/10.1103/PhysRevA.101.053413} {\bibfield  {journal} {\bibinfo
   {journal} {Phys. Rev. A}\ }\textbf {\bibinfo {volume} {101}},\ \bibinfo
  {pages} {053413} (\bibinfo {year} {2020})}\BibitemShut {NoStop}%
\bibitem [{\citenamefont {Jensen}\ \emph {et~al.}(2016)\citenamefont {Jensen},
  \citenamefont {Fl{\aa}}, \citenamefont {Jonsson}, \citenamefont {Monstad},
  \citenamefont {Ruud},\ and\ \citenamefont {Frediani}}]{jensen2016magnetic}%
  \BibitemOpen
  \bibfield  {author} {\bibinfo {author} {\bibfnamefont {S.~R.}\ \bibnamefont
  {Jensen}}, \bibinfo {author} {\bibfnamefont {T.}~\bibnamefont {Fl{\aa}}},
  \bibinfo {author} {\bibfnamefont {D.}~\bibnamefont {Jonsson}}, \bibinfo
  {author} {\bibfnamefont {R.~S.}\ \bibnamefont {Monstad}}, \bibinfo {author}
  {\bibfnamefont {K.}~\bibnamefont {Ruud}},\ and\ \bibinfo {author}
  {\bibfnamefont {L.}~\bibnamefont {Frediani}},\ }\bibfield  {title} {\bibinfo
  {title} {Magnetic properties with multiwavelets and dft: the complete basis
  set limit achieved},\ }\href {https://doi.org/10.1039/C6CP01294A} {\bibfield
  {journal} {\bibinfo  {journal} {Physical Chemistry Chemical Physics}\
  }\textbf {\bibinfo {volume} {18}},\ \bibinfo {pages} {21145} (\bibinfo {year}
  {2016})}\BibitemShut {NoStop}%
\bibitem [{\citenamefont {Anderson}\ \emph {et~al.}(2020)\citenamefont
  {Anderson}, \citenamefont {Harrison}, \citenamefont {Sundahl}, \citenamefont
  {Thornton},\ and\ \citenamefont {Beylkin}}]{ANDERSON2020112711}%
  \BibitemOpen
  \bibfield  {author} {\bibinfo {author} {\bibfnamefont {J.}~\bibnamefont
  {Anderson}}, \bibinfo {author} {\bibfnamefont {R.~J.}\ \bibnamefont
  {Harrison}}, \bibinfo {author} {\bibfnamefont {B.}~\bibnamefont {Sundahl}},
  \bibinfo {author} {\bibfnamefont {W.~S.}\ \bibnamefont {Thornton}},\ and\
  \bibinfo {author} {\bibfnamefont {G.}~\bibnamefont {Beylkin}},\ }\bibfield
  {title} {\bibinfo {title} {Real-space quasi-relativistic quantum chemistry},\
  }\href {https://doi.org/https://doi.org/10.1016/j.comptc.2020.112711}
  {\bibfield  {journal} {\bibinfo  {journal} {Computational and Theoretical
  Chemistry}\ }\textbf {\bibinfo {volume} {1175}},\ \bibinfo {pages} {112711}
  (\bibinfo {year} {2020})}\BibitemShut {NoStop}%
\bibitem [{\citenamefont {Anderson}\ \emph {et~al.}(2019)\citenamefont
  {Anderson}, \citenamefont {Sundahl}, \citenamefont {Harrison},\ and\
  \citenamefont {Beylkin}}]{anderson2019dirac}%
  \BibitemOpen
  \bibfield  {author} {\bibinfo {author} {\bibfnamefont {J.}~\bibnamefont
  {Anderson}}, \bibinfo {author} {\bibfnamefont {B.}~\bibnamefont {Sundahl}},
  \bibinfo {author} {\bibfnamefont {R.}~\bibnamefont {Harrison}},\ and\
  \bibinfo {author} {\bibfnamefont {G.}~\bibnamefont {Beylkin}},\ }\bibfield
  {title} {\bibinfo {title} {Dirac-fock calculations on molecules in an
  adaptive multiwavelet basis},\ }\href {https://doi.org/10.1063/1.5128908}
  {\bibfield  {journal} {\bibinfo  {journal} {The Journal of Chemical Physics}\
  }\textbf {\bibinfo {volume} {151}},\ \bibinfo {pages} {234112} (\bibinfo
  {year} {2019})},\ \Eprint
  {https://arxiv.org/abs/https://doi.org/10.1063/1.5128908}
  {https://doi.org/10.1063/1.5128908} \BibitemShut {NoStop}%
\bibitem [{\citenamefont {Bischoff}\ \emph {et~al.}(2012)\citenamefont
  {Bischoff}, \citenamefont {Harrison},\ and\ \citenamefont
  {Valeev}}]{bischoff2012}%
  \BibitemOpen
  \bibfield  {author} {\bibinfo {author} {\bibfnamefont {F.~A.}\ \bibnamefont
  {Bischoff}}, \bibinfo {author} {\bibfnamefont {R.~J.}\ \bibnamefont
  {Harrison}},\ and\ \bibinfo {author} {\bibfnamefont {E.~F.}\ \bibnamefont
  {Valeev}},\ }\bibfield  {title} {\bibinfo {title} {Computing many-body wave
  functions with guaranteed precision: The first-order møller-plesset wave
  function for the ground state of helium atom},\ }\href
  {https://doi.org/10.1063/1.4747538} {\bibfield  {journal} {\bibinfo
  {journal} {The Journal of Chemical Physics}\ }\textbf {\bibinfo {volume}
  {137}},\ \bibinfo {pages} {104103} (\bibinfo {year} {2012})},\ \Eprint
  {https://arxiv.org/abs/https://doi.org/10.1063/1.4747538}
  {https://doi.org/10.1063/1.4747538} \BibitemShut {NoStop}%
\bibitem [{\citenamefont {Bischoff}\ and\ \citenamefont
  {Valeev}(2013)}]{bischoff2013}%
  \BibitemOpen
  \bibfield  {author} {\bibinfo {author} {\bibfnamefont {F.~A.}\ \bibnamefont
  {Bischoff}}\ and\ \bibinfo {author} {\bibfnamefont {E.~F.}\ \bibnamefont
  {Valeev}},\ }\bibfield  {title} {\bibinfo {title} {Computing molecular
  correlation energies with guaranteed precision},\ }\href
  {https://doi.org/10.1063/1.4820404} {\bibfield  {journal} {\bibinfo
  {journal} {The Journal of Chemical Physics}\ }\textbf {\bibinfo {volume}
  {139}},\ \bibinfo {pages} {114106} (\bibinfo {year} {2013})},\ \Eprint
  {https://arxiv.org/abs/https://doi.org/10.1063/1.4820404}
  {https://doi.org/10.1063/1.4820404} \BibitemShut {NoStop}%
\bibitem [{\citenamefont {Kottmann}\ and\ \citenamefont
  {Bischoff}(2017{\natexlab{a}})}]{kottmann2017coupledGS}%
  \BibitemOpen
  \bibfield  {author} {\bibinfo {author} {\bibfnamefont {J.~S.}\ \bibnamefont
  {Kottmann}}\ and\ \bibinfo {author} {\bibfnamefont {F.~A.}\ \bibnamefont
  {Bischoff}},\ }\bibfield  {title} {\bibinfo {title} {Coupled-cluster in real
  space. {1. CC2} ground state energies using multiresolution analysis},\
  }\href {https://doi.org/10.1021/acs.jctc.7b00694} {\bibfield  {journal}
  {\bibinfo  {journal} {Journal of chemical theory and computation}\ }\textbf
  {\bibinfo {volume} {13}},\ \bibinfo {pages} {5945} (\bibinfo {year}
  {2017}{\natexlab{a}})}\BibitemShut {NoStop}%
\bibitem [{\citenamefont {Kottmann}\ and\ \citenamefont
  {Bischoff}(2017{\natexlab{b}})}]{kottmann2017coupledES}%
  \BibitemOpen
  \bibfield  {author} {\bibinfo {author} {\bibfnamefont {J.~S.}\ \bibnamefont
  {Kottmann}}\ and\ \bibinfo {author} {\bibfnamefont {F.~A.}\ \bibnamefont
  {Bischoff}},\ }\bibfield  {title} {\bibinfo {title} {Coupled-cluster in real
  space. {2. CC2} excited states using multiresolution analysis},\ }\href
  {https://doi.org/10.1021/acs.jctc.7b00695} {\bibfield  {journal} {\bibinfo
  {journal} {Journal of chemical theory and computation}\ }\textbf {\bibinfo
  {volume} {13}},\ \bibinfo {pages} {5956} (\bibinfo {year}
  {2017}{\natexlab{b}})}\BibitemShut {NoStop}%
\bibitem [{\citenamefont {Sosa}\ \emph {et~al.}(1989)\citenamefont {Sosa},
  \citenamefont {Geertsen}, \citenamefont {Trucks}, \citenamefont {Bartlett},\
  and\ \citenamefont {Franz}}]{sosa1989selection}%
  \BibitemOpen
  \bibfield  {author} {\bibinfo {author} {\bibfnamefont {C.}~\bibnamefont
  {Sosa}}, \bibinfo {author} {\bibfnamefont {J.}~\bibnamefont {Geertsen}},
  \bibinfo {author} {\bibfnamefont {G.~W.}\ \bibnamefont {Trucks}}, \bibinfo
  {author} {\bibfnamefont {R.~J.}\ \bibnamefont {Bartlett}},\ and\ \bibinfo
  {author} {\bibfnamefont {J.~A.}\ \bibnamefont {Franz}},\ }\bibfield  {title}
  {\bibinfo {title} {Selection of the reduced virtual space for correlated
  calculations. an application to the energy and dipole moment of h2o},\
  }\href@noop {} {\bibfield  {journal} {\bibinfo  {journal} {Chemical physics
  letters}\ }\textbf {\bibinfo {volume} {159}},\ \bibinfo {pages} {148}
  (\bibinfo {year} {1989})}\BibitemShut {NoStop}%
\bibitem [{\citenamefont {DePrince~III}\ and\ \citenamefont
  {Sherrill}(2013)}]{deprince2013accurate}%
  \BibitemOpen
  \bibfield  {author} {\bibinfo {author} {\bibfnamefont {A.~E.}\ \bibnamefont
  {DePrince~III}}\ and\ \bibinfo {author} {\bibfnamefont {C.~D.}\ \bibnamefont
  {Sherrill}},\ }\bibfield  {title} {\bibinfo {title} {Accurate noncovalent
  interaction energies using truncated basis sets based on frozen natural
  orbitals},\ }\href@noop {} {\bibfield  {journal} {\bibinfo  {journal}
  {Journal of chemical theory and computation}\ }\textbf {\bibinfo {volume}
  {9}},\ \bibinfo {pages} {293} (\bibinfo {year} {2013})}\BibitemShut {NoStop}%
\bibitem [{\citenamefont {Riplinger}\ and\ \citenamefont
  {Neese}(2013)}]{riplinger2013}%
  \BibitemOpen
  \bibfield  {author} {\bibinfo {author} {\bibfnamefont {C.}~\bibnamefont
  {Riplinger}}\ and\ \bibinfo {author} {\bibfnamefont {F.}~\bibnamefont
  {Neese}},\ }\bibfield  {title} {\bibinfo {title} {An efficient and near
  linear scaling pair natural orbital based local coupled cluster method},\
  }\href {https://doi.org/10.1063/1.4773581} {\bibfield  {journal} {\bibinfo
  {journal} {The Journal of Chemical Physics}\ }\textbf {\bibinfo {volume}
  {138}},\ \bibinfo {pages} {034106} (\bibinfo {year} {2013})},\ \Eprint
  {https://arxiv.org/abs/https://doi.org/10.1063/1.4773581}
  {https://doi.org/10.1063/1.4773581} \BibitemShut {NoStop}%
\bibitem [{\citenamefont {Pinski}\ \emph {et~al.}(2015)\citenamefont {Pinski},
  \citenamefont {Riplinger}, \citenamefont {Valeev},\ and\ \citenamefont
  {Neese}}]{Pinski:2015ii}%
  \BibitemOpen
  \bibfield  {author} {\bibinfo {author} {\bibfnamefont {P.}~\bibnamefont
  {Pinski}}, \bibinfo {author} {\bibfnamefont {C.}~\bibnamefont {Riplinger}},
  \bibinfo {author} {\bibfnamefont {E.~F.}\ \bibnamefont {Valeev}},\ and\
  \bibinfo {author} {\bibfnamefont {F.}~\bibnamefont {Neese}},\ }\bibfield
  {title} {\bibinfo {title} {{Sparse maps---A systematic infrastructure for
  reduced-scaling electronic structure methods. I. An efficient and simple
  linear scaling local MP2 method that uses an intermediate basis of pair
  natural orbitals}},\ }\href@noop {} {\bibfield  {journal} {\bibinfo
  {journal} {J Chem Phys}\ }\textbf {\bibinfo {volume} {143}},\ \bibinfo
  {pages} {034108} (\bibinfo {year} {2015})}\BibitemShut {NoStop}%
\bibitem [{\citenamefont {Babbush}\ \emph {et~al.}(2019)\citenamefont
  {Babbush}, \citenamefont {Berry}, \citenamefont {McClean},\ and\
  \citenamefont {Neven}}]{babbush2019quantum}%
  \BibitemOpen
  \bibfield  {author} {\bibinfo {author} {\bibfnamefont {R.}~\bibnamefont
  {Babbush}}, \bibinfo {author} {\bibfnamefont {D.~W.}\ \bibnamefont {Berry}},
  \bibinfo {author} {\bibfnamefont {J.~R.}\ \bibnamefont {McClean}},\ and\
  \bibinfo {author} {\bibfnamefont {H.}~\bibnamefont {Neven}},\ }\bibfield
  {title} {\bibinfo {title} {Quantum simulation of chemistry with sublinear
  scaling in basis size},\ }\href {https://doi.org/10.1038/s41534-019-0199-y}
  {\bibfield  {journal} {\bibinfo  {journal} {npj Quantum Information}\
  }\textbf {\bibinfo {volume} {5}},\ \bibinfo {pages} {1} (\bibinfo {year}
  {2019})}\BibitemShut {NoStop}%
\bibitem [{\citenamefont {McClean}\ \emph {et~al.}(2020)\citenamefont
  {McClean}, \citenamefont {Faulstich}, \citenamefont {Zhu}, \citenamefont
  {O'Gorman}, \citenamefont {Qiu}, \citenamefont {White}, \citenamefont
  {Babbush},\ and\ \citenamefont {Lin}}]{mcclean2020gausslet}%
  \BibitemOpen
  \bibfield  {author} {\bibinfo {author} {\bibfnamefont {J.}~\bibnamefont
  {McClean}}, \bibinfo {author} {\bibfnamefont {F.}~\bibnamefont {Faulstich}},
  \bibinfo {author} {\bibfnamefont {Q.}~\bibnamefont {Zhu}}, \bibinfo {author}
  {\bibfnamefont {B.}~\bibnamefont {O'Gorman}}, \bibinfo {author}
  {\bibfnamefont {Y.}~\bibnamefont {Qiu}}, \bibinfo {author} {\bibfnamefont
  {S.~R.}\ \bibnamefont {White}}, \bibinfo {author} {\bibfnamefont
  {R.}~\bibnamefont {Babbush}},\ and\ \bibinfo {author} {\bibfnamefont
  {L.}~\bibnamefont {Lin}},\ }\bibfield  {title} {\bibinfo {title}
  {Discontinuous galerkin discretization for quantum simulation of chemistry},\
  }\href {http://iopscience.iop.org/10.1088/1367-2630/ab9d9f} {\bibfield
  {journal} {\bibinfo  {journal} {New Journal of Physics}\ } (\bibinfo {year}
  {2020})}\BibitemShut {NoStop}%
\bibitem [{\citenamefont {Peruzzo}\ \emph {et~al.}(2014)\citenamefont
  {Peruzzo}, \citenamefont {McClean}, \citenamefont {Shadbolt}, \citenamefont
  {Yung}, \citenamefont {Zhou}, \citenamefont {Love}, \citenamefont
  {Aspuru-Guzik},\ and\ \citenamefont {O’brien}}]{peruzzo2014variational}%
  \BibitemOpen
  \bibfield  {author} {\bibinfo {author} {\bibfnamefont {A.}~\bibnamefont
  {Peruzzo}}, \bibinfo {author} {\bibfnamefont {J.}~\bibnamefont {McClean}},
  \bibinfo {author} {\bibfnamefont {P.}~\bibnamefont {Shadbolt}}, \bibinfo
  {author} {\bibfnamefont {M.-H.}\ \bibnamefont {Yung}}, \bibinfo {author}
  {\bibfnamefont {X.-Q.}\ \bibnamefont {Zhou}}, \bibinfo {author}
  {\bibfnamefont {P.~J.}\ \bibnamefont {Love}}, \bibinfo {author}
  {\bibfnamefont {A.}~\bibnamefont {Aspuru-Guzik}},\ and\ \bibinfo {author}
  {\bibfnamefont {J.~L.}\ \bibnamefont {O’brien}},\ }\bibfield  {title}
  {\bibinfo {title} {A variational eigenvalue solver on a photonic quantum
  processor},\ }\href {https://doi.org/10.1038/ncomms5213} {\bibfield
  {journal} {\bibinfo  {journal} {Nature communications}\ }\textbf {\bibinfo
  {volume} {5}},\ \bibinfo {pages} {4213} (\bibinfo {year} {2014})}\BibitemShut
  {NoStop}%
\bibitem [{\citenamefont {McClean}\ \emph {et~al.}(2016)\citenamefont
  {McClean}, \citenamefont {Romero}, \citenamefont {Babbush},\ and\
  \citenamefont {Aspuru-Guzik}}]{McClean2016theoryofvqe}%
  \BibitemOpen
  \bibfield  {author} {\bibinfo {author} {\bibfnamefont {J.~R.}\ \bibnamefont
  {McClean}}, \bibinfo {author} {\bibfnamefont {J.}~\bibnamefont {Romero}},
  \bibinfo {author} {\bibfnamefont {R.}~\bibnamefont {Babbush}},\ and\ \bibinfo
  {author} {\bibfnamefont {A.}~\bibnamefont {Aspuru-Guzik}},\ }\bibfield
  {title} {\bibinfo {title} {The theory of variational hybrid quantum-classical
  algorithms},\ }\href {https://doi.org/10.1088/1367-2630/18/2/023023}
  {\bibfield  {journal} {\bibinfo  {journal} {New Journal of Physics}\ }\textbf
  {\bibinfo {volume} {18}},\ \bibinfo {pages} {023023} (\bibinfo {year}
  {2016})}\BibitemShut {NoStop}%
\bibitem [{\citenamefont {Kottmann}\ \emph
  {et~al.}(2020{\natexlab{a}})\citenamefont {Kottmann}, \citenamefont
  {Bischoff},\ and\ \citenamefont {Valeev}}]{kottmann2020direct}%
  \BibitemOpen
  \bibfield  {author} {\bibinfo {author} {\bibfnamefont {J.~S.}\ \bibnamefont
  {Kottmann}}, \bibinfo {author} {\bibfnamefont {F.~A.}\ \bibnamefont
  {Bischoff}},\ and\ \bibinfo {author} {\bibfnamefont {E.~F.}\ \bibnamefont
  {Valeev}},\ }\bibfield  {title} {\bibinfo {title} {Direct determination of
  optimal pair-natural orbitals in a real-space representation: The
  second-order moller--plesset energy},\ }\href
  {https://doi.org/10.1063/1.5141880} {\bibfield  {journal} {\bibinfo
  {journal} {The Journal of Chemical Physics}\ }\textbf {\bibinfo {volume}
  {152}},\ \bibinfo {pages} {074105} (\bibinfo {year}
  {2020}{\natexlab{a}})}\BibitemShut {NoStop}%
\bibitem [{\citenamefont {Takeshita}\ \emph {et~al.}(2020)\citenamefont
  {Takeshita}, \citenamefont {Rubin}, \citenamefont {Jiang}, \citenamefont
  {Lee}, \citenamefont {Babbush},\ and\ \citenamefont
  {McClean}}]{takeshita_increasing_2020}%
  \BibitemOpen
  \bibfield  {author} {\bibinfo {author} {\bibfnamefont {T.}~\bibnamefont
  {Takeshita}}, \bibinfo {author} {\bibfnamefont {N.~C.}\ \bibnamefont
  {Rubin}}, \bibinfo {author} {\bibfnamefont {Z.}~\bibnamefont {Jiang}},
  \bibinfo {author} {\bibfnamefont {E.}~\bibnamefont {Lee}}, \bibinfo {author}
  {\bibfnamefont {R.}~\bibnamefont {Babbush}},\ and\ \bibinfo {author}
  {\bibfnamefont {J.~R.}\ \bibnamefont {McClean}},\ }\bibfield  {title}
  {\bibinfo {title} {Increasing the {Representation} {Accuracy} of {Quantum}
  {Simulations} of {Chemistry} without {Extra} {Quantum} {Resources}},\ }\href
  {https://doi.org/10.1103/PhysRevX.10.011004} {\bibfield  {journal} {\bibinfo
  {journal} {Physical Review X}\ }\textbf {\bibinfo {volume} {10}},\ \bibinfo
  {pages} {011004} (\bibinfo {year} {2020})},\ \bibinfo {note} {publisher:
  American Physical Society}\BibitemShut {NoStop}%
\bibitem [{\citenamefont {Motta}\ \emph {et~al.}(2020)\citenamefont {Motta},
  \citenamefont {Gujarati}, \citenamefont {Rice}, \citenamefont {Kumar},
  \citenamefont {Masteran}, \citenamefont {Latone}, \citenamefont {Lee},
  \citenamefont {Valeev},\ and\ \citenamefont
  {Takeshita}}]{motta_quantum_2020}%
  \BibitemOpen
  \bibfield  {author} {\bibinfo {author} {\bibfnamefont {M.}~\bibnamefont
  {Motta}}, \bibinfo {author} {\bibfnamefont {T.~P.}\ \bibnamefont {Gujarati}},
  \bibinfo {author} {\bibfnamefont {J.~E.}\ \bibnamefont {Rice}}, \bibinfo
  {author} {\bibfnamefont {A.}~\bibnamefont {Kumar}}, \bibinfo {author}
  {\bibfnamefont {C.}~\bibnamefont {Masteran}}, \bibinfo {author}
  {\bibfnamefont {J.~A.}\ \bibnamefont {Latone}}, \bibinfo {author}
  {\bibfnamefont {E.}~\bibnamefont {Lee}}, \bibinfo {author} {\bibfnamefont
  {E.~F.}\ \bibnamefont {Valeev}},\ and\ \bibinfo {author} {\bibfnamefont
  {T.~Y.}\ \bibnamefont {Takeshita}},\ }\bibfield  {title} {\bibinfo {title}
  {Quantum simulation of electronic structure with transcorrelated
  {Hamiltonian}: increasing accuracy without extra quantum resources},\ }\href
  {http://arxiv.org/abs/2006.02488} {\bibfield  {journal} {\bibinfo  {journal}
  {arXiv:2006.02488 [physics, physics:quant-ph]}\ } (\bibinfo {year} {2020})},\
  \bibinfo {note} {arXiv: 2006.02488}\BibitemShut {NoStop}%
\bibitem [{\citenamefont {McArdle}\ and\ \citenamefont
  {Tew}(2020)}]{mcardle2020improving}%
  \BibitemOpen
  \bibfield  {author} {\bibinfo {author} {\bibfnamefont {S.}~\bibnamefont
  {McArdle}}\ and\ \bibinfo {author} {\bibfnamefont {D.~P.}\ \bibnamefont
  {Tew}},\ }\bibfield  {title} {\bibinfo {title} {Improving the accuracy of
  quantum computational chemistry using the transcorrelated method},\ }\href
  {https://arxiv.org/abs/2006.11181} {\bibfield  {journal} {\bibinfo  {journal}
  {arXiv preprint arXiv:2006.11181}\ } (\bibinfo {year} {2020})}\BibitemShut
  {NoStop}%
\bibitem [{\citenamefont {Shavitt}\ and\ \citenamefont
  {Bartlett}(2009)}]{shavitt2009many}%
  \BibitemOpen
  \bibfield  {author} {\bibinfo {author} {\bibfnamefont {I.}~\bibnamefont
  {Shavitt}}\ and\ \bibinfo {author} {\bibfnamefont {R.~J.}\ \bibnamefont
  {Bartlett}},\ }\href@noop {} {\emph {\bibinfo {title} {Many-body methods in
  chemistry and physics: MBPT and coupled-cluster theory}}}\ (\bibinfo
  {publisher} {Cambridge university press},\ \bibinfo {year}
  {2009})\BibitemShut {NoStop}%
\bibitem [{\citenamefont {J{\o}rgensen}(2012)}]{jorgensen2012second}%
  \BibitemOpen
  \bibfield  {author} {\bibinfo {author} {\bibfnamefont {P.}~\bibnamefont
  {J{\o}rgensen}},\ }\href@noop {} {\emph {\bibinfo {title} {Second
  quantization-based methods in quantum chemistry}}}\ (\bibinfo  {publisher}
  {Elsevier},\ \bibinfo {year} {2012})\BibitemShut {NoStop}%
\bibitem [{\citenamefont {Surj{\'a}n}(2012)}]{surjan2012second}%
  \BibitemOpen
  \bibfield  {author} {\bibinfo {author} {\bibfnamefont {P.~R.}\ \bibnamefont
  {Surj{\'a}n}},\ }\href@noop {} {\emph {\bibinfo {title} {Second quantized
  approach to quantum chemistry: an elementary introduction}}}\ (\bibinfo
  {publisher} {Springer Science \& Business Media},\ \bibinfo {year}
  {2012})\BibitemShut {NoStop}%
\bibitem [{\citenamefont {Jordan}\ and\ \citenamefont
  {Klein}(1927)}]{jordan1927mehrkorperproblem}%
  \BibitemOpen
  \bibfield  {author} {\bibinfo {author} {\bibfnamefont {P.}~\bibnamefont
  {Jordan}}\ and\ \bibinfo {author} {\bibfnamefont {O.}~\bibnamefont {Klein}},\
  }\bibfield  {title} {\bibinfo {title} {Zum mehrk{\"o}rperproblem der
  quantentheorie},\ }\href@noop {} {\bibfield  {journal} {\bibinfo  {journal}
  {Zeitschrift f{\"u}r Physik}\ }\textbf {\bibinfo {volume} {45}},\ \bibinfo
  {pages} {751} (\bibinfo {year} {1927})}\BibitemShut {NoStop}%
\bibitem [{\citenamefont {Bravyi}\ and\ \citenamefont
  {Kitaev}(2002)}]{bravyi2002}%
  \BibitemOpen
  \bibfield  {author} {\bibinfo {author} {\bibfnamefont {S.~B.}\ \bibnamefont
  {Bravyi}}\ and\ \bibinfo {author} {\bibfnamefont {A.~Y.}\ \bibnamefont
  {Kitaev}},\ }\bibfield  {title} {\bibinfo {title} {Fermionic quantum
  computation},\ }\href
  {https://doi.org/https://doi.org/10.1006/aphy.2002.6254} {\bibfield
  {journal} {\bibinfo  {journal} {Annals of Physics}\ }\textbf {\bibinfo
  {volume} {298}},\ \bibinfo {pages} {210 } (\bibinfo {year}
  {2002})}\BibitemShut {NoStop}%
\bibitem [{\citenamefont {Seeley}\ \emph {et~al.}(2012)\citenamefont {Seeley},
  \citenamefont {Richard},\ and\ \citenamefont {Love}}]{seeley2012}%
  \BibitemOpen
  \bibfield  {author} {\bibinfo {author} {\bibfnamefont {J.~T.}\ \bibnamefont
  {Seeley}}, \bibinfo {author} {\bibfnamefont {M.~J.}\ \bibnamefont
  {Richard}},\ and\ \bibinfo {author} {\bibfnamefont {P.~J.}\ \bibnamefont
  {Love}},\ }\bibfield  {title} {\bibinfo {title} {The bravyi-kitaev
  transformation for quantum computation of electronic structure},\ }\href
  {https://doi.org/10.1063/1.4768229} {\bibfield  {journal} {\bibinfo
  {journal} {The Journal of Chemical Physics}\ }\textbf {\bibinfo {volume}
  {137}},\ \bibinfo {pages} {224109} (\bibinfo {year} {2012})},\ \Eprint
  {https://arxiv.org/abs/https://doi.org/10.1063/1.4768229}
  {https://doi.org/10.1063/1.4768229} \BibitemShut {NoStop}%
\bibitem [{\citenamefont {Setia}\ and\ \citenamefont
  {Whitfield}(2018)}]{setia2018}%
  \BibitemOpen
  \bibfield  {author} {\bibinfo {author} {\bibfnamefont {K.}~\bibnamefont
  {Setia}}\ and\ \bibinfo {author} {\bibfnamefont {J.~D.}\ \bibnamefont
  {Whitfield}},\ }\bibfield  {title} {\bibinfo {title} {Bravyi-kitaev superfast
  simulation of electronic structure on a quantum computer},\ }\href
  {https://doi.org/10.1063/1.5019371} {\bibfield  {journal} {\bibinfo
  {journal} {The Journal of Chemical Physics}\ }\textbf {\bibinfo {volume}
  {148}},\ \bibinfo {pages} {164104} (\bibinfo {year} {2018})},\ \Eprint
  {https://arxiv.org/abs/https://doi.org/10.1063/1.5019371}
  {https://doi.org/10.1063/1.5019371} \BibitemShut {NoStop}%
\bibitem [{\citenamefont {Barison}\ \emph {et~al.}(2020)\citenamefont
  {Barison}, \citenamefont {Galli},\ and\ \citenamefont
  {Motta}}]{barison2020quantum}%
  \BibitemOpen
  \bibfield  {author} {\bibinfo {author} {\bibfnamefont {S.}~\bibnamefont
  {Barison}}, \bibinfo {author} {\bibfnamefont {D.~E.}\ \bibnamefont {Galli}},\
  and\ \bibinfo {author} {\bibfnamefont {M.}~\bibnamefont {Motta}},\
  }\href@noop {} {\bibinfo {title} {Quantum simulations of molecular systems
  with intrinsic atomic orbitals}} (\bibinfo {year} {2020}),\ \Eprint
  {https://arxiv.org/abs/2011.08137} {arXiv:2011.08137 [quant-ph]} \BibitemShut
  {NoStop}%
\bibitem [{\citenamefont {Kalos}(1962)}]{Kalos1962}%
  \BibitemOpen
  \bibfield  {author} {\bibinfo {author} {\bibfnamefont {M.~H.}\ \bibnamefont
  {Kalos}},\ }\bibfield  {title} {\bibinfo {title} {Monte carlo calculations of
  the ground state of three- and four-body nuclei},\ }\href
  {https://doi.org/10.1103/PhysRev.128.1791} {\bibfield  {journal} {\bibinfo
  {journal} {Phys. Rev.}\ }\textbf {\bibinfo {volume} {128}},\ \bibinfo {pages}
  {1791} (\bibinfo {year} {1962})}\BibitemShut {NoStop}%
\bibitem [{\citenamefont {Beylkin}\ and\ \citenamefont
  {Mohlenkamp}(2005)}]{beylkin2005alg}%
  \BibitemOpen
  \bibfield  {author} {\bibinfo {author} {\bibfnamefont {G.}~\bibnamefont
  {Beylkin}}\ and\ \bibinfo {author} {\bibfnamefont {M.~J.}\ \bibnamefont
  {Mohlenkamp}},\ }\bibfield  {title} {\bibinfo {title} {Algorithms for
  numerical analysis in high dimensions},\ }\href
  {https://doi.org/10.1137/040604959} {\bibfield  {journal} {\bibinfo
  {journal} {SIAM Journal on Scientific Computing}\ }\textbf {\bibinfo {volume}
  {26}},\ \bibinfo {pages} {2133} (\bibinfo {year} {2005})},\ \Eprint
  {https://arxiv.org/abs/https://doi.org/10.1137/040604959}
  {https://doi.org/10.1137/040604959} \BibitemShut {NoStop}%
\bibitem [{\citenamefont {Kottmann}(2018)}]{Kottmann2018Coupled-Cluster}%
  \BibitemOpen
  \bibfield  {author} {\bibinfo {author} {\bibfnamefont {J.~S.}\ \bibnamefont
  {Kottmann}},\ }\emph {\bibinfo {title} {Coupled-Cluster in Real Space}},\
  \href {https://doi.org/http://dx.doi.org/10.18452/19357} {Ph.D. thesis},\
  \bibinfo  {school} {Humboldt-Universität zu Berlin,
  Mathematisch-Naturwissenschaftliche Fakultät} (\bibinfo {year}
  {2018})\BibitemShut {NoStop}%
\bibitem [{\citenamefont {Neese}\ \emph {et~al.}(2009)\citenamefont {Neese},
  \citenamefont {Hansen},\ and\ \citenamefont {Liakos}}]{Neese:2009db}%
  \BibitemOpen
  \bibfield  {author} {\bibinfo {author} {\bibfnamefont {F.}~\bibnamefont
  {Neese}}, \bibinfo {author} {\bibfnamefont {A.}~\bibnamefont {Hansen}},\ and\
  \bibinfo {author} {\bibfnamefont {D.~G.}\ \bibnamefont {Liakos}},\ }\bibfield
   {title} {\bibinfo {title} {{Efficient and accurate approximations to the
  local coupled cluster singles doubles method using a truncated pair natural
  orbital basis}},\ }\href@noop {} {\bibfield  {journal} {\bibinfo  {journal}
  {J Chem Phys}\ }\textbf {\bibinfo {volume} {131}},\ \bibinfo {pages} {064103}
  (\bibinfo {year} {2009})}\BibitemShut {NoStop}%
\bibitem [{\citenamefont {Werner}\ \emph {et~al.}(2015)\citenamefont {Werner},
  \citenamefont {Knizia}, \citenamefont {Krause}, \citenamefont {Schwilk},\
  and\ \citenamefont {Dornbach}}]{Werner:2015kw}%
  \BibitemOpen
  \bibfield  {author} {\bibinfo {author} {\bibfnamefont {H.-J.}\ \bibnamefont
  {Werner}}, \bibinfo {author} {\bibfnamefont {G.}~\bibnamefont {Knizia}},
  \bibinfo {author} {\bibfnamefont {C.}~\bibnamefont {Krause}}, \bibinfo
  {author} {\bibfnamefont {M.}~\bibnamefont {Schwilk}},\ and\ \bibinfo {author}
  {\bibfnamefont {M.}~\bibnamefont {Dornbach}},\ }\bibfield  {title} {\bibinfo
  {title} {{Scalable Electron Correlation Methods I.: PNO-LMP2 with Linear
  Scaling in the Molecular Size and Near-Inverse-Linear Scaling in the Number
  of Processors}},\ }\href@noop {} {\bibfield  {journal} {\bibinfo  {journal}
  {J. Chem. Theory Comput.}\ }\textbf {\bibinfo {volume} {11}},\ \bibinfo
  {pages} {484} (\bibinfo {year} {2015})}\BibitemShut {NoStop}%
\bibitem [{\citenamefont {Yanai}\ \emph {et~al.}(2004)\citenamefont {Yanai},
  \citenamefont {Fann}, \citenamefont {Gan}, \citenamefont {Harrison},\ and\
  \citenamefont {Beylkin}}]{yanai2004multiresolution}%
  \BibitemOpen
  \bibfield  {author} {\bibinfo {author} {\bibfnamefont {T.}~\bibnamefont
  {Yanai}}, \bibinfo {author} {\bibfnamefont {G.~I.}\ \bibnamefont {Fann}},
  \bibinfo {author} {\bibfnamefont {Z.}~\bibnamefont {Gan}}, \bibinfo {author}
  {\bibfnamefont {R.~J.}\ \bibnamefont {Harrison}},\ and\ \bibinfo {author}
  {\bibfnamefont {G.}~\bibnamefont {Beylkin}},\ }\bibfield  {title} {\bibinfo
  {title} {Multiresolution quantum chemistry in multiwavelet bases:
  Hartree–fock exchange},\ }\href {https://doi.org/10.1063/1.1790931}
  {\bibfield  {journal} {\bibinfo  {journal} {The Journal of Chemical Physics}\
  }\textbf {\bibinfo {volume} {121}},\ \bibinfo {pages} {6680} (\bibinfo {year}
  {2004})},\ \Eprint {https://arxiv.org/abs/https://doi.org/10.1063/1.1790931}
  {https://doi.org/10.1063/1.1790931} \BibitemShut {NoStop}%
\bibitem [{\citenamefont {Elfving}\ \emph {et~al.}(2020)\citenamefont
  {Elfving}, \citenamefont {Gámez},\ and\ \citenamefont
  {Gogolin}}]{elfving2020simulating}%
  \BibitemOpen
  \bibfield  {author} {\bibinfo {author} {\bibfnamefont {V.~E.}\ \bibnamefont
  {Elfving}}, \bibinfo {author} {\bibfnamefont {J.~A.}\ \bibnamefont
  {Gámez}},\ and\ \bibinfo {author} {\bibfnamefont {C.}~\bibnamefont
  {Gogolin}},\ }\href@noop {} {\bibinfo {title} {Simulating quantum chemistry
  in the restricted hartree-fock space on a qubit-based quantum computing
  device}} (\bibinfo {year} {2020}),\ \Eprint
  {https://arxiv.org/abs/2002.00035} {arXiv:2002.00035 [quant-ph]} \BibitemShut
  {NoStop}%
\bibitem [{\citenamefont {Khamoshi}\ \emph {et~al.}(2020)\citenamefont
  {Khamoshi}, \citenamefont {Evangelista},\ and\ \citenamefont
  {Scuseria}}]{khamoshi2020correlating}%
  \BibitemOpen
  \bibfield  {author} {\bibinfo {author} {\bibfnamefont {A.}~\bibnamefont
  {Khamoshi}}, \bibinfo {author} {\bibfnamefont {F.~A.}\ \bibnamefont
  {Evangelista}},\ and\ \bibinfo {author} {\bibfnamefont {G.~E.}\ \bibnamefont
  {Scuseria}},\ }\href@noop {} {\bibinfo {title} {Correlating agp on a quantum
  computer}} (\bibinfo {year} {2020}),\ \Eprint
  {https://arxiv.org/abs/2008.06138} {arXiv:2008.06138 [quant-ph]} \BibitemShut
  {NoStop}%
\bibitem [{\citenamefont {Lee}\ \emph {et~al.}(2018)\citenamefont {Lee},
  \citenamefont {Huggins}, \citenamefont {Head-Gordon},\ and\ \citenamefont
  {Whaley}}]{lee2018generalized}%
  \BibitemOpen
  \bibfield  {author} {\bibinfo {author} {\bibfnamefont {J.}~\bibnamefont
  {Lee}}, \bibinfo {author} {\bibfnamefont {W.~J.}\ \bibnamefont {Huggins}},
  \bibinfo {author} {\bibfnamefont {M.}~\bibnamefont {Head-Gordon}},\ and\
  \bibinfo {author} {\bibfnamefont {K.~B.}\ \bibnamefont {Whaley}},\ }\bibfield
   {title} {\bibinfo {title} {Generalized unitary coupled cluster wave
  functions for quantum computation},\ }\href
  {https://doi.org/10.1021/acs.jctc.8b01004} {\bibfield  {journal} {\bibinfo
  {journal} {Journal of chemical theory and computation}\ }\textbf {\bibinfo
  {volume} {15}},\ \bibinfo {pages} {311} (\bibinfo {year} {2018})}\BibitemShut
  {NoStop}%
\bibitem [{\citenamefont {Kottmann}\ \emph
  {et~al.}(2020{\natexlab{b}})\citenamefont {Kottmann}, \citenamefont
  {Alperin-Lea}, \citenamefont {Tamayo-Mendoza}, \citenamefont
  {Cervera-Lierta}, \citenamefont {Lavigne}, \citenamefont {Yen}, \citenamefont
  {Verteletskyi}, \citenamefont {Schleich}, \citenamefont {Anand},
  \citenamefont {Degroote}, \citenamefont {Chaney}, \citenamefont {Kesibi},
  \citenamefont {Curnow}, \citenamefont {Izmaylov},\ and\ \citenamefont
  {Aspuru-Guzik}}]{tequila}%
  \BibitemOpen
  \bibfield  {author} {\bibinfo {author} {\bibfnamefont {J.~S.}\ \bibnamefont
  {Kottmann}}, \bibinfo {author} {\bibfnamefont {S.}~\bibnamefont
  {Alperin-Lea}}, \bibinfo {author} {\bibfnamefont {T.}~\bibnamefont
  {Tamayo-Mendoza}}, \bibinfo {author} {\bibfnamefont {A.}~\bibnamefont
  {Cervera-Lierta}}, \bibinfo {author} {\bibfnamefont {C.}~\bibnamefont
  {Lavigne}}, \bibinfo {author} {\bibfnamefont {T.-C.}\ \bibnamefont {Yen}},
  \bibinfo {author} {\bibfnamefont {V.}~\bibnamefont {Verteletskyi}}, \bibinfo
  {author} {\bibfnamefont {P.}~\bibnamefont {Schleich}}, \bibinfo {author}
  {\bibfnamefont {A.}~\bibnamefont {Anand}}, \bibinfo {author} {\bibfnamefont
  {M.}~\bibnamefont {Degroote}}, \bibinfo {author} {\bibfnamefont
  {S.}~\bibnamefont {Chaney}}, \bibinfo {author} {\bibfnamefont
  {M.}~\bibnamefont {Kesibi}}, \bibinfo {author} {\bibfnamefont {N.~G.}\
  \bibnamefont {Curnow}}, \bibinfo {author} {\bibfnamefont {A.~F.}\
  \bibnamefont {Izmaylov}},\ and\ \bibinfo {author} {\bibfnamefont
  {A.}~\bibnamefont {Aspuru-Guzik}},\ }\href {https://arxiv.org/abs/2011.03057}
  {\bibinfo {title} {Tequila: A platform for rapid development of quantum
  algorithms}} (\bibinfo {year} {2020}{\natexlab{b}}),\ \Eprint
  {https://arxiv.org/abs/2011.03057} {2011.03057 [quant-ph]} \BibitemShut
  {NoStop}%
\bibitem [{\citenamefont {Suzuki}\ \emph {et~al.}(2020)\citenamefont {Suzuki},
  \citenamefont {Kawase}, \citenamefont {Masumura}, \citenamefont {Hiraga},
  \citenamefont {Nakadai}, \citenamefont {Chen}, \citenamefont {Nakanishi},
  \citenamefont {Mitarai}, \citenamefont {Imai}, \citenamefont {Tamiya},
  \citenamefont {Yamamoto}, \citenamefont {Yan}, \citenamefont {Kawakubo},
  \citenamefont {Nakagawa}, \citenamefont {Ibe}, \citenamefont {Zhang},
  \citenamefont {Yamashita}, \citenamefont {Yoshimura}, \citenamefont
  {Hayashi},\ and\ \citenamefont {Fujii}}]{qulacs}%
  \BibitemOpen
  \bibfield  {author} {\bibinfo {author} {\bibfnamefont {Y.}~\bibnamefont
  {Suzuki}}, \bibinfo {author} {\bibfnamefont {Y.}~\bibnamefont {Kawase}},
  \bibinfo {author} {\bibfnamefont {Y.}~\bibnamefont {Masumura}}, \bibinfo
  {author} {\bibfnamefont {Y.}~\bibnamefont {Hiraga}}, \bibinfo {author}
  {\bibfnamefont {M.}~\bibnamefont {Nakadai}}, \bibinfo {author} {\bibfnamefont
  {J.}~\bibnamefont {Chen}}, \bibinfo {author} {\bibfnamefont {K.~M.}\
  \bibnamefont {Nakanishi}}, \bibinfo {author} {\bibfnamefont {K.}~\bibnamefont
  {Mitarai}}, \bibinfo {author} {\bibfnamefont {R.}~\bibnamefont {Imai}},
  \bibinfo {author} {\bibfnamefont {S.}~\bibnamefont {Tamiya}}, \bibinfo
  {author} {\bibfnamefont {T.}~\bibnamefont {Yamamoto}}, \bibinfo {author}
  {\bibfnamefont {T.}~\bibnamefont {Yan}}, \bibinfo {author} {\bibfnamefont
  {T.}~\bibnamefont {Kawakubo}}, \bibinfo {author} {\bibfnamefont {Y.~O.}\
  \bibnamefont {Nakagawa}}, \bibinfo {author} {\bibfnamefont {Y.}~\bibnamefont
  {Ibe}}, \bibinfo {author} {\bibfnamefont {Y.}~\bibnamefont {Zhang}}, \bibinfo
  {author} {\bibfnamefont {H.}~\bibnamefont {Yamashita}}, \bibinfo {author}
  {\bibfnamefont {H.}~\bibnamefont {Yoshimura}}, \bibinfo {author}
  {\bibfnamefont {A.}~\bibnamefont {Hayashi}},\ and\ \bibinfo {author}
  {\bibfnamefont {K.}~\bibnamefont {Fujii}},\ }\href
  {https://arxiv.org/abs/2011.13524} {\bibinfo {title} {Qulacs: a fast and
  versatile quantum circuit simulator for research purpose}} (\bibinfo {year}
  {2020}),\ \Eprint {https://arxiv.org/abs/2011.13524} {2011.13524 [quant-ph]}
  \BibitemShut {NoStop}%
\bibitem [{\citenamefont {{P. Virtanen, R. Gommers \textit{et. al.
  }}}(2020)}]{scipy}%
  \BibitemOpen
  \bibfield  {author} {\bibinfo {author} {\bibnamefont {{P. Virtanen, R.
  Gommers \textit{et. al. }}}},\ }\bibfield  {title} {\bibinfo {title} {{SciPy
  1.0: Fundamental Algorithms for Scientific Computing in Python}},\ }\href
  {https://doi.org/https://doi.org/10.1038/s41592-019-0686-2} {\bibfield
  {journal} {\bibinfo  {journal} {Nature Methods}\ }\textbf {\bibinfo {volume}
  {17}},\ \bibinfo {pages} {261} (\bibinfo {year} {2020})}\BibitemShut
  {NoStop}%
\bibitem [{\citenamefont {{J.R. McClean \textit{et.
  al.}}}(2017)}]{openfermion}%
  \BibitemOpen
  \bibfield  {author} {\bibinfo {author} {\bibnamefont {{J.R. McClean
  \textit{et. al.}}}},\ }\href {https://arxiv.org/abs/1710.07629} {\bibinfo
  {title} {Openfermion: The electronic structure package for quantum
  computers}} (\bibinfo {year} {2017}),\ \Eprint
  {https://arxiv.org/abs/1710.07629} {arXiv:1710.07629:1710.07629 [quant-ph]}
  \BibitemShut {NoStop}%
\bibitem [{\citenamefont {Kottmann}\ \emph
  {et~al.}(2020{\natexlab{c}})\citenamefont {Kottmann}, \citenamefont {Anand},\
  and\ \citenamefont {Aspuru-Guzik}}]{kottmann2020feasible}%
  \BibitemOpen
  \bibfield  {author} {\bibinfo {author} {\bibfnamefont {J.~S.}\ \bibnamefont
  {Kottmann}}, \bibinfo {author} {\bibfnamefont {A.}~\bibnamefont {Anand}},\
  and\ \bibinfo {author} {\bibfnamefont {A.}~\bibnamefont {Aspuru-Guzik}},\
  }\href {https://arxiv.org/abs/2011.05938} {\bibinfo {title} {A feasible
  approach for automatically differentiable unitary coupled-cluster on quantum
  computers}} (\bibinfo {year} {2020}{\natexlab{c}}),\ \Eprint
  {https://arxiv.org/abs/2011.05938} {2011.05938 [quant-ph]} \BibitemShut
  {NoStop}%
\bibitem [{\citenamefont {Smith}\ \emph {et~al.}(2020)\citenamefont {Smith},
  \citenamefont {Burns}, \citenamefont {Simmonett}, \citenamefont {Parrish},
  \citenamefont {Schieber}, \citenamefont {Galvelis}, \citenamefont {Kraus},
  \citenamefont {Kruse}, \citenamefont {Di~Remigio}, \citenamefont {Alenaizan},
  \citenamefont {James}, \citenamefont {Lehtola}, \citenamefont {Misiewicz},
  \citenamefont {Scheurer}, \citenamefont {Shaw}, \citenamefont {Schriber},
  \citenamefont {Xie}, \citenamefont {Glick}, \citenamefont {Sirianni},
  \citenamefont {O’Brien}, \citenamefont {Waldrop}, \citenamefont {Kumar},
  \citenamefont {Hohenstein}, \citenamefont {Pritchard}, \citenamefont
  {Brooks}, \citenamefont {Schaefer}, \citenamefont {Sokolov}, \citenamefont
  {Patkowski}, \citenamefont {DePrince}, \citenamefont {Bozkaya}, \citenamefont
  {King}, \citenamefont {Evangelista}, \citenamefont {Turney}, \citenamefont
  {Crawford},\ and\ \citenamefont {Sherrill}}]{psi42020}%
  \BibitemOpen
  \bibfield  {author} {\bibinfo {author} {\bibfnamefont {D.~G.~A.}\
  \bibnamefont {Smith}}, \bibinfo {author} {\bibfnamefont {L.~A.}\ \bibnamefont
  {Burns}}, \bibinfo {author} {\bibfnamefont {A.~C.}\ \bibnamefont
  {Simmonett}}, \bibinfo {author} {\bibfnamefont {R.~M.}\ \bibnamefont
  {Parrish}}, \bibinfo {author} {\bibfnamefont {M.~C.}\ \bibnamefont
  {Schieber}}, \bibinfo {author} {\bibfnamefont {R.}~\bibnamefont {Galvelis}},
  \bibinfo {author} {\bibfnamefont {P.}~\bibnamefont {Kraus}}, \bibinfo
  {author} {\bibfnamefont {H.}~\bibnamefont {Kruse}}, \bibinfo {author}
  {\bibfnamefont {R.}~\bibnamefont {Di~Remigio}}, \bibinfo {author}
  {\bibfnamefont {A.}~\bibnamefont {Alenaizan}}, \bibinfo {author}
  {\bibfnamefont {A.~M.}\ \bibnamefont {James}}, \bibinfo {author}
  {\bibfnamefont {S.}~\bibnamefont {Lehtola}}, \bibinfo {author} {\bibfnamefont
  {J.~P.}\ \bibnamefont {Misiewicz}}, \bibinfo {author} {\bibfnamefont
  {M.}~\bibnamefont {Scheurer}}, \bibinfo {author} {\bibfnamefont {R.~A.}\
  \bibnamefont {Shaw}}, \bibinfo {author} {\bibfnamefont {J.~B.}\ \bibnamefont
  {Schriber}}, \bibinfo {author} {\bibfnamefont {Y.}~\bibnamefont {Xie}},
  \bibinfo {author} {\bibfnamefont {Z.~L.}\ \bibnamefont {Glick}}, \bibinfo
  {author} {\bibfnamefont {D.~A.}\ \bibnamefont {Sirianni}}, \bibinfo {author}
  {\bibfnamefont {J.~S.}\ \bibnamefont {O’Brien}}, \bibinfo {author}
  {\bibfnamefont {J.~M.}\ \bibnamefont {Waldrop}}, \bibinfo {author}
  {\bibfnamefont {A.}~\bibnamefont {Kumar}}, \bibinfo {author} {\bibfnamefont
  {E.~G.}\ \bibnamefont {Hohenstein}}, \bibinfo {author} {\bibfnamefont
  {B.~P.}\ \bibnamefont {Pritchard}}, \bibinfo {author} {\bibfnamefont {B.~R.}\
  \bibnamefont {Brooks}}, \bibinfo {author} {\bibfnamefont {H.~F.}\
  \bibnamefont {Schaefer}}, \bibinfo {author} {\bibfnamefont {A.~Y.}\
  \bibnamefont {Sokolov}}, \bibinfo {author} {\bibfnamefont {K.}~\bibnamefont
  {Patkowski}}, \bibinfo {author} {\bibfnamefont {A.~E.}\ \bibnamefont
  {DePrince}}, \bibinfo {author} {\bibfnamefont {U.}~\bibnamefont {Bozkaya}},
  \bibinfo {author} {\bibfnamefont {R.~A.}\ \bibnamefont {King}}, \bibinfo
  {author} {\bibfnamefont {F.~A.}\ \bibnamefont {Evangelista}}, \bibinfo
  {author} {\bibfnamefont {J.~M.}\ \bibnamefont {Turney}}, \bibinfo {author}
  {\bibfnamefont {T.~D.}\ \bibnamefont {Crawford}},\ and\ \bibinfo {author}
  {\bibfnamefont {C.~D.}\ \bibnamefont {Sherrill}},\ }\bibfield  {title}
  {\bibinfo {title} {Psi4 1.4: Open-source software for high-throughput quantum
  chemistry},\ }\href {https://doi.org/10.1063/5.0006002} {\bibfield  {journal}
  {\bibinfo  {journal} {The Journal of Chemical Physics}\ }\textbf {\bibinfo
  {volume} {152}},\ \bibinfo {pages} {184108} (\bibinfo {year}
  {2020})}\BibitemShut {NoStop}%
\bibitem [{\citenamefont {Bravyi}\ \emph {et~al.}(2017)\citenamefont {Bravyi},
  \citenamefont {Gambetta}, \citenamefont {Mezzacapo},\ and\ \citenamefont
  {Temme}}]{bravyi2017tapering}%
  \BibitemOpen
  \bibfield  {author} {\bibinfo {author} {\bibfnamefont {S.}~\bibnamefont
  {Bravyi}}, \bibinfo {author} {\bibfnamefont {J.~M.}\ \bibnamefont
  {Gambetta}}, \bibinfo {author} {\bibfnamefont {A.}~\bibnamefont
  {Mezzacapo}},\ and\ \bibinfo {author} {\bibfnamefont {K.}~\bibnamefont
  {Temme}},\ }\bibfield  {title} {\bibinfo {title} {Tapering off qubits to
  simulate fermionic hamiltonians},\ }\href@noop {} {\bibfield  {journal}
  {\bibinfo  {journal} {arXiv preprint arXiv:1701.08213}\ } (\bibinfo {year}
  {2017})}\BibitemShut {NoStop}%
\bibitem [{\citenamefont {Dehesa}\ \emph {et~al.}(1992)\citenamefont {Dehesa},
  \citenamefont {Angulo}, \citenamefont {Koga},\ and\ \citenamefont
  {Matsui}}]{dehesa_study_1992}%
  \BibitemOpen
  \bibfield  {author} {\bibinfo {author} {\bibfnamefont {J.~S.}\ \bibnamefont
  {Dehesa}}, \bibinfo {author} {\bibfnamefont {J.~C.}\ \bibnamefont {Angulo}},
  \bibinfo {author} {\bibfnamefont {T.}~\bibnamefont {Koga}},\ and\ \bibinfo
  {author} {\bibfnamefont {K.}~\bibnamefont {Matsui}},\ }\bibfield  {title}
  {{\selectlanguage {english}\bibinfo {title} {Study of some interelectronic
  properties in helium-like atoms}},\ }\href
  {https://doi.org/10.1007/BF01437514} {\bibfield  {journal} {\bibinfo
  {journal} {Zeitschrift für Physik D Atoms, Molecules and Clusters}\ }\textbf
  {\bibinfo {volume} {25}},\ \bibinfo {pages} {9} (\bibinfo {year}
  {1992})}\BibitemShut {NoStop}%
\bibitem [{\citenamefont {Nakatsuji}(2012)}]{hiroshi2012}%
  \BibitemOpen
  \bibfield  {author} {\bibinfo {author} {\bibfnamefont {H.}~\bibnamefont
  {Nakatsuji}},\ }\bibfield  {title} {\bibinfo {title} {Discovery of a general
  method of solving the schrödinger and dirac equations that opens a way to
  accurately predictive quantum chemistry},\ }\href
  {https://doi.org/10.1021/ar200340j} {\bibfield  {journal} {\bibinfo
  {journal} {Accounts of Chemical Research}\ }\textbf {\bibinfo {volume}
  {45}},\ \bibinfo {pages} {1480} (\bibinfo {year} {2012})},\ \bibinfo {note}
  {pMID: 22686372},\ \Eprint
  {https://arxiv.org/abs/https://doi.org/10.1021/ar200340j}
  {https://doi.org/10.1021/ar200340j} \BibitemShut {NoStop}%
\bibitem [{\citenamefont
  {Bischoff}(2014{\natexlab{a}})}]{bischoff2014regularizingmany}%
  \BibitemOpen
  \bibfield  {author} {\bibinfo {author} {\bibfnamefont {F.~A.}\ \bibnamefont
  {Bischoff}},\ }\bibfield  {title} {\bibinfo {title} {Regularizing the
  molecular potential in electronic structure calculations. {II. Many-body}
  methods},\ }\href {https://doi.org/10.1063/1.4901022} {\bibfield  {journal}
  {\bibinfo  {journal} {The Journal of Chemical Physics}\ }\textbf {\bibinfo
  {volume} {141}},\ \bibinfo {pages} {184106} (\bibinfo {year}
  {2014}{\natexlab{a}})},\ \Eprint
  {https://arxiv.org/abs/https://doi.org/10.1063/1.4901022}
  {https://doi.org/10.1063/1.4901022} \BibitemShut {NoStop}%
\bibitem [{\citenamefont {Kato}(1957)}]{Kato1957cusp}%
  \BibitemOpen
  \bibfield  {author} {\bibinfo {author} {\bibfnamefont {T.}~\bibnamefont
  {Kato}},\ }\bibfield  {title} {\bibinfo {title} {On the eigenfunctions of
  many-particle systems in quantum mechanics},\ }\href
  {https://doi.org/10.1002/cpa.3160100201} {\bibfield  {journal} {\bibinfo
  {journal} {Communications on Pure and Applied Mathematics}\ }\textbf
  {\bibinfo {volume} {10}},\ \bibinfo {pages} {151} (\bibinfo {year}
  {1957})}\BibitemShut {NoStop}%
\bibitem [{\citenamefont
  {Bischoff}(2014{\natexlab{b}})}]{bischoff2014regularizing}%
  \BibitemOpen
  \bibfield  {author} {\bibinfo {author} {\bibfnamefont {F.~A.}\ \bibnamefont
  {Bischoff}},\ }\bibfield  {title} {\bibinfo {title} {Regularizing the
  molecular potential in electronic structure calculations. {I. SCF} methods},\
  }\href {https://doi.org/10.1063/1.4901021} {\bibfield  {journal} {\bibinfo
  {journal} {The Journal of chemical physics}\ }\textbf {\bibinfo {volume}
  {141}},\ \bibinfo {pages} {184105} (\bibinfo {year}
  {2014}{\natexlab{b}})}\BibitemShut {NoStop}%
\bibitem [{\citenamefont {Kutzelnigg}(1985)}]{kutzelnigg1985r}%
  \BibitemOpen
  \bibfield  {author} {\bibinfo {author} {\bibfnamefont {W.}~\bibnamefont
  {Kutzelnigg}},\ }\bibfield  {title} {\bibinfo {title} {r12-dependent terms in
  the wave function as closed sums of partial wave amplitudes for large l},\
  }\href {https://doi.org/10.1007/BF00527669} {\bibfield  {journal} {\bibinfo
  {journal} {Theoretica chimica acta}\ }\textbf {\bibinfo {volume} {68}},\
  \bibinfo {pages} {445} (\bibinfo {year} {1985})}\BibitemShut {NoStop}%
\bibitem [{\citenamefont {Tamayo-Mendoza}\ \emph {et~al.}(2018)\citenamefont
  {Tamayo-Mendoza}, \citenamefont {Kreisbeck}, \citenamefont {Lindh},\ and\
  \citenamefont {Aspuru-Guzik}}]{tamayo2018}%
  \BibitemOpen
  \bibfield  {author} {\bibinfo {author} {\bibfnamefont {T.}~\bibnamefont
  {Tamayo-Mendoza}}, \bibinfo {author} {\bibfnamefont {C.}~\bibnamefont
  {Kreisbeck}}, \bibinfo {author} {\bibfnamefont {R.}~\bibnamefont {Lindh}},\
  and\ \bibinfo {author} {\bibfnamefont {A.}~\bibnamefont {Aspuru-Guzik}},\
  }\bibfield  {title} {\bibinfo {title} {Automatic differentiation in quantum
  chemistry with applications to fully variational hartree–fock},\ }\href
  {https://doi.org/10.1021/acscentsci.7b00586} {\bibfield  {journal} {\bibinfo
  {journal} {ACS Central Science}\ }\textbf {\bibinfo {volume} {4}},\ \bibinfo
  {pages} {559} (\bibinfo {year} {2018})},\ \Eprint
  {https://arxiv.org/abs/https://doi.org/10.1021/acscentsci.7b00586}
  {https://doi.org/10.1021/acscentsci.7b00586} \BibitemShut {NoStop}%
\bibitem [{\citenamefont {Bande}\ \emph {et~al.}(2010)\citenamefont {Bande},
  \citenamefont {Nakashima},\ and\ \citenamefont {Nakatsuji}}]{bande_lih_2010}%
  \BibitemOpen
  \bibfield  {author} {\bibinfo {author} {\bibfnamefont {A.}~\bibnamefont
  {Bande}}, \bibinfo {author} {\bibfnamefont {H.}~\bibnamefont {Nakashima}},\
  and\ \bibinfo {author} {\bibfnamefont {H.}~\bibnamefont {Nakatsuji}},\
  }\bibfield  {title} {{\selectlanguage {english}\bibinfo {title} {{LiH}
  potential energy curves for ground and excited states with the free
  complement local {Schrödinger} equation method}},\ }\href
  {https://doi.org/10.1016/j.cplett.2010.07.041} {\bibfield  {journal}
  {\bibinfo  {journal} {Chemical Physics Letters}\ }\textbf {\bibinfo {volume}
  {496}},\ \bibinfo {pages} {347} (\bibinfo {year} {2010})}\BibitemShut
  {NoStop}%
\bibitem [{\citenamefont {{{T. Tamayo-Mendoza \textit{et.
  al.}}}}(2018)}]{diffiqult}%
  \BibitemOpen
  \bibfield  {author} {\bibinfo {author} {\bibnamefont {{{T. Tamayo-Mendoza
  \textit{et. al.}}}}},\ }\href
  {https://github.com/aspuru-guzik-group/DiffiQult.git} {\bibinfo {title}
  {{DiffiQult}}} (\bibinfo {year} {2018})\BibitemShut {NoStop}%
\bibitem [{\citenamefont {Grimsley}\ \emph {et~al.}(2019)\citenamefont
  {Grimsley}, \citenamefont {Economou}, \citenamefont {Barnes},\ and\
  \citenamefont {Mayhall}}]{grimsley2019adaptive}%
  \BibitemOpen
  \bibfield  {author} {\bibinfo {author} {\bibfnamefont {H.~R.}\ \bibnamefont
  {Grimsley}}, \bibinfo {author} {\bibfnamefont {S.~E.}\ \bibnamefont
  {Economou}}, \bibinfo {author} {\bibfnamefont {E.}~\bibnamefont {Barnes}},\
  and\ \bibinfo {author} {\bibfnamefont {N.~J.}\ \bibnamefont {Mayhall}},\
  }\bibfield  {title} {\bibinfo {title} {An adaptive variational algorithm for
  exact molecular simulations on a quantum computer},\ }\href
  {https://doi.org/10.1038/s41467-019-10988-2} {\bibfield  {journal} {\bibinfo
  {journal} {Nature communications}\ }\textbf {\bibinfo {volume} {10}},\
  \bibinfo {pages} {1} (\bibinfo {year} {2019})}\BibitemShut {NoStop}%
\bibitem [{\citenamefont {Sokolov}\ \emph {et~al.}(2020)\citenamefont
  {Sokolov}, \citenamefont {Barkoutsos}, \citenamefont {Ollitrault},
  \citenamefont {Greenberg}, \citenamefont {Rice}, \citenamefont {Pistoia},\
  and\ \citenamefont {Tavernelli}}]{sokolov2020quantum}%
  \BibitemOpen
  \bibfield  {author} {\bibinfo {author} {\bibfnamefont {I.~O.}\ \bibnamefont
  {Sokolov}}, \bibinfo {author} {\bibfnamefont {P.~K.}\ \bibnamefont
  {Barkoutsos}}, \bibinfo {author} {\bibfnamefont {P.~J.}\ \bibnamefont
  {Ollitrault}}, \bibinfo {author} {\bibfnamefont {D.}~\bibnamefont
  {Greenberg}}, \bibinfo {author} {\bibfnamefont {J.}~\bibnamefont {Rice}},
  \bibinfo {author} {\bibfnamefont {M.}~\bibnamefont {Pistoia}},\ and\ \bibinfo
  {author} {\bibfnamefont {I.}~\bibnamefont {Tavernelli}},\ }\bibfield  {title}
  {\bibinfo {title} {Quantum orbital-optimized unitary coupled cluster methods
  in the strongly correlated regime: Can quantum algorithms outperform their
  classical equivalents?},\ }\href {https://doi.org/10.1063/1.5141835}
  {\bibfield  {journal} {\bibinfo  {journal} {The Journal of Chemical Physics}\
  }\textbf {\bibinfo {volume} {152}},\ \bibinfo {pages} {124107} (\bibinfo
  {year} {2020})},\ \Eprint
  {https://arxiv.org/abs/https://doi.org/10.1063/1.5141835}
  {https://doi.org/10.1063/1.5141835} \BibitemShut {NoStop}%
\bibitem [{\citenamefont {Ryabinkin}\ \emph {et~al.}(2018)\citenamefont
  {Ryabinkin}, \citenamefont {Yen}, \citenamefont {Genin},\ and\ \citenamefont
  {Izmaylov}}]{ryabinkin2018qubit}%
  \BibitemOpen
  \bibfield  {author} {\bibinfo {author} {\bibfnamefont {I.~G.}\ \bibnamefont
  {Ryabinkin}}, \bibinfo {author} {\bibfnamefont {T.-C.}\ \bibnamefont {Yen}},
  \bibinfo {author} {\bibfnamefont {S.~N.}\ \bibnamefont {Genin}},\ and\
  \bibinfo {author} {\bibfnamefont {A.~F.}\ \bibnamefont {Izmaylov}},\
  }\bibfield  {title} {\bibinfo {title} {Qubit coupled cluster method: a
  systematic approach to quantum chemistry on a quantum computer},\ }\href
  {https://doi.org/10.1021/acs.jctc.8b00932} {\bibfield  {journal} {\bibinfo
  {journal} {Journal of chemical theory and computation}\ }\textbf {\bibinfo
  {volume} {14}},\ \bibinfo {pages} {6317} (\bibinfo {year}
  {2018})}\BibitemShut {NoStop}%
\bibitem [{\citenamefont {Setia}\ \emph {et~al.}(2019)\citenamefont {Setia},
  \citenamefont {Chen}, \citenamefont {Rice}, \citenamefont {Mezzacapo},
  \citenamefont {Pistoia},\ and\ \citenamefont
  {Whitfield}}]{setia2019reducing}%
  \BibitemOpen
  \bibfield  {author} {\bibinfo {author} {\bibfnamefont {K.}~\bibnamefont
  {Setia}}, \bibinfo {author} {\bibfnamefont {R.}~\bibnamefont {Chen}},
  \bibinfo {author} {\bibfnamefont {J.~E.}\ \bibnamefont {Rice}}, \bibinfo
  {author} {\bibfnamefont {A.}~\bibnamefont {Mezzacapo}}, \bibinfo {author}
  {\bibfnamefont {M.}~\bibnamefont {Pistoia}},\ and\ \bibinfo {author}
  {\bibfnamefont {J.}~\bibnamefont {Whitfield}},\ }\bibfield  {title} {\bibinfo
  {title} {Reducing qubit requirements for quantum simulation using molecular
  point group symmetries},\ }\href {https://arxiv.org/abs/1910.14644}
  {\bibfield  {journal} {\bibinfo  {journal} {arXiv preprint arXiv:1910.14644}\
  } (\bibinfo {year} {2019})}\BibitemShut {NoStop}%
\bibitem [{\citenamefont {Ponce}\ \emph {et~al.}(2019)\citenamefont {Ponce},
  \citenamefont {van Zon}, \citenamefont {Northrup}, \citenamefont {Gruner},
  \citenamefont {Chen}, \citenamefont {Ertinaz}, \citenamefont {Fedoseev},
  \citenamefont {Groer}, \citenamefont {Mao}, \citenamefont {Mundim} \emph
  {et~al.}}]{niagara1}%
  \BibitemOpen
  \bibfield  {author} {\bibinfo {author} {\bibfnamefont {M.}~\bibnamefont
  {Ponce}}, \bibinfo {author} {\bibfnamefont {R.}~\bibnamefont {van Zon}},
  \bibinfo {author} {\bibfnamefont {S.}~\bibnamefont {Northrup}}, \bibinfo
  {author} {\bibfnamefont {D.}~\bibnamefont {Gruner}}, \bibinfo {author}
  {\bibfnamefont {J.}~\bibnamefont {Chen}}, \bibinfo {author} {\bibfnamefont
  {F.}~\bibnamefont {Ertinaz}}, \bibinfo {author} {\bibfnamefont
  {A.}~\bibnamefont {Fedoseev}}, \bibinfo {author} {\bibfnamefont
  {L.}~\bibnamefont {Groer}}, \bibinfo {author} {\bibfnamefont
  {F.}~\bibnamefont {Mao}}, \bibinfo {author} {\bibfnamefont {B.~C.}\
  \bibnamefont {Mundim}}, \emph {et~al.},\ }\bibfield  {title} {\bibinfo
  {title} {Deploying a top-100 supercomputer for large parallel workloads: The
  niagara supercomputer},\ }in\ \href@noop {} {\emph {\bibinfo {booktitle}
  {Proceedings of the Practice and Experience in Advanced Research Computing on
  Rise of the Machines (learning)}}}\ (\bibinfo {year} {2019})\ pp.\ \bibinfo
  {pages} {1--8}\BibitemShut {NoStop}%
\bibitem [{\citenamefont {Loken}\ \emph {et~al.}(2010)\citenamefont {Loken},
  \citenamefont {Gruner}, \citenamefont {Groer}, \citenamefont {Peltier},
  \citenamefont {Bunn}, \citenamefont {Craig}, \citenamefont {Henriques},
  \citenamefont {Dempsey}, \citenamefont {Yu}, \citenamefont {Chen} \emph
  {et~al.}}]{niagara2}%
  \BibitemOpen
  \bibfield  {author} {\bibinfo {author} {\bibfnamefont {C.}~\bibnamefont
  {Loken}}, \bibinfo {author} {\bibfnamefont {D.}~\bibnamefont {Gruner}},
  \bibinfo {author} {\bibfnamefont {L.}~\bibnamefont {Groer}}, \bibinfo
  {author} {\bibfnamefont {R.}~\bibnamefont {Peltier}}, \bibinfo {author}
  {\bibfnamefont {N.}~\bibnamefont {Bunn}}, \bibinfo {author} {\bibfnamefont
  {M.}~\bibnamefont {Craig}}, \bibinfo {author} {\bibfnamefont
  {T.}~\bibnamefont {Henriques}}, \bibinfo {author} {\bibfnamefont
  {J.}~\bibnamefont {Dempsey}}, \bibinfo {author} {\bibfnamefont {C.-H.}\
  \bibnamefont {Yu}}, \bibinfo {author} {\bibfnamefont {J.}~\bibnamefont
  {Chen}}, \emph {et~al.},\ }\bibfield  {title} {\bibinfo {title} {Scinet:
  lessons learned from building a power-efficient top-20 system and data
  centre},\ }in\ \href@noop {} {\emph {\bibinfo {booktitle} {Journal of
  Physics-Conference Series}}},\ Vol.\ \bibinfo {volume} {256}\ (\bibinfo
  {year} {2010})\ p.\ \bibinfo {pages} {012026}\BibitemShut {NoStop}%
\end{thebibliography}%

\end{document}